\documentclass[a4paper,USenglish,thm-restate,cleveref]{lipics-v2021}
\hideLIPIcs
\nolinenumbers 

\usepackage[ruled]{algorithm}
\usepackage[noend]{algpseudocode} 
\usepackage{cite}

\algblock{Phase}{EndPhase}
\algblockdefx[Phase]{START}{}%
    [1][If]{\textbf{#1}}%
\algnotext[Phase]{EndPhase}

\newcommand{\valueRead}{\mathit{val}}
\newcommand{\valueAudit}{\mathit{audit\_response}}

\newcommand{\temp}{\mathit{val}}
\newcommand{\auditResult}{\mathit{audit\_result}}
\newcommand{\readResult}{\mathit{read\_result}}

\newcommand{\currVal}{\mathit{curr\_val}}
\newcommand{\prevVal}{\mathit{prev\_val}}

\newcommand{\tempWriteMr}{\mathit{val}}
\newcommand{\bit}{\mathit{bit}}
\newcommand{\tempMr}{\mathit{val}}

\newcommand{\Nreader}{\mathit{n}}

\newcommand{\Values}{\mathit{values}}
\newcommand{\AuditResponse}{\mathit{audit\_response}}
\newcommand{\SafeValues}{\mathit{safe\_values}}

\newcommand{\swmrmaIwrite}{{write}}
\newcommand{\swmrmaIaudit}{{audit}}

\newcommand{\readLogRegMA}{\mathit{read\_log}}
\newcommand{\readValueRegMA}{\mathit{read\_result}}

\newcommand{\op}[1]{\mathit{op_{#1}}}

\def\BibTeX{{\rm B\kern-.05em{\sc i\kern-.025em b}\kern-.08em
    T\kern-.1667em\lower.7ex\hbox{E}\kern-.125emX}}

\title{The Synchronization Power of Auditable Registers}

\author{Hagit Attiya}{Technion, Israel}{ hagit@cs.technion.ac.il}{https://orcid.org/}{}
\author{Antonella Del Pozzo}{Université Paris-Saclay, CEA, List, F-91120, Palaiseau, France }{antonella.delpozzo@cea.fr}{https://orcid.org/}{}
\author{Alessia Milani}{Aix-Marseille Université, France}{alessia.milani@lis.fr}{https://orcid.org/}{}
\author{Ulysse Pavloff}{Université Paris-Saclay, CEA, List, F-91120, Palaiseau, France }{ulysse.pavloff@cea.fr}{https://orcid.org/}{}
\author{Alexandre Rapetti}{Université Paris-Saclay, CEA, List, F-91120, Palaiseau, France }{alexandre.rapetti@cea.fr}{https://orcid.org/0009-0008-3151-6495}{}

\authorrunning{Attiya, Del Pozzo, Milani, Pavloff and Rapetti} 



\begin{document}


\maketitle

\begin{abstract}
Auditability allows to track all the read operations performed on a register. 
It abstracts the need of data owners to control access to their data,
tracking who read which information. 
This work considers possible formalizations of auditing and their ramification 
for the possibility of providing it. 

The natural definition is to require a linearization of all write, read and audit 
operations together (\emph{atomic} auditing). 
The paper shows that atomic auditing is a powerful tool,
\emph{as it can be used to solve consensus}.
The number of processes that can solve consensus using atomic audit 
depends on the number of processes that can read or audit the register.
If there is a single reader or a single auditor (the writer), 
then consensus can be solved among two processes.
If multiple readers and auditors are possible, 
then consensus can be solved among the same number of processes.
This means that strong synchronization primitives are needed to support 
atomic auditing. 

We give implementations of atomic audit when there are either multiple 
readers or multiple auditors (but not both) 
using primitives with consensus number 2 (swap and fetch\&add).
When there are multiple readers \emph{and} multiple auditors,
the implementation uses compare\&swap.

These findings motivate a weaker definition, in which 
audit operations are not linearized together with the write and read operations
(\emph{regular} auditing).
We prove that regular auditing can be implemented from ordinary reads 
and writes on atomic registers. 
\end{abstract}


\section{Introduction}
Outsourcing {storage} capabilities to third-party distributed 
storage is a common practice for both private and professional users. 
It helps to circumvent local space constraints, dependability, 
and accessibility limitations. 
Unfortunately, this means having to trust the distributed storage 
provider on data integrity, retrievability, and confidentiality. 
Those issues are underscored by relentless attacks on data storage 
servers~\cite{DB18}, 
which increased the awareness to data confidentiality and sovereignty
and lead to a recent worldwide deployment of data protection 
regulations~\cite{gdpr,ccpa,pipl}. 
Even in secure storage systems where \emph{access control} policies 
regulate who can access the data, an unauthorized user can access data either due to a misconfiguration 
of the access control system or in the occurrence of a data breach~\cite{a2be09,ABESurvey20}.

As a result, data owners are increasingly concerned about who accesses their data. 
%
This makes \emph{auditability}---the systematic tracking of 
who has read data in storage systems and the information it has 
observed---an important feature.

A \emph{register} is an abstraction for distributed 
storage that provides read and write operations to the clients. 
An \emph{auditable register}, introduced by Cogo and Bessani~\cite{CB21},
is a register enriched with an \emph{audit} operation.
An audit operation indicates who performed read operations on it, 
and which values they have read.
Auditability is defined in terms of two properties: 
\emph{completeness} ensures that readers that access data are detected,
while \emph{accuracy} ensures that readers who do not access data are not incriminated. 

In this work, we formalize the correctness of auditable registers in terms of their 
high-level operations (read, write and audit). 
We first formalize a natural extension of an atomic register, called \emph{atomic register with atomic audit}
where all the operations (including audit) appear to happen in a sequential order that respects their real-time order.

We show that an \emph{atomic register with atomic audit} is a powerful abstraction, 
\emph{because it has a greater consensus number than an ordinary atomic register}. 
Recall that the \emph{consensus number}~\cite{H91} of object type $X$ 
is $m$ if $m$ is the largest integer such that 
there exists an asynchronous consensus algorithm for $m$ processes,
up to $m-1$ may crash, using only shared objects of type $X$ 
and read/write registers. 

We present a wait-free algorithm that solves consensus 
among \emph{two processes}, using an atomic register with atomic audit 
where only one process (the writer) can perform audit operations. 
This stands in contrast to the well-known result~\cite{H91} 
that atomic read/write registers cannot be used to solve wait-free 
consensus among two processes. 
We then show that when $m$ processes can read and audit the register,
it is possible to solve consensus among $m$ processes.

These results indicate that base objects stronger than read/write registers
are needed to implement atomic audit,
motivating our implementations of an atomic register with atomic audit. 
When there is either a single auditor or a single reader, 
we use base objects with consensus number 2
($\mathit{swap}$ and $\mathit{fetch\&add}$).

Specifically, we first present a simple algorithm for a 
single-reader atomic auditable register with atomic audit, 
where the writer is the only process that can execute the audit operations.
The writer needs to atomically retrieve who read the previously written 
value while writing a new value. 
With a single reader, 
this can be easily ensured by using $\mathit{swap}$ primitives: 
to read a value, the reader atomically swaps it with a special value. 
If the writer retrieves this special value when writing a new value, 
then it is aware of the value read. 

Extending this idea to multi-readers is challenging,
since readers might be swapping each other's values. 
We propose a solution that uses a single shared object accessed 
with $\mathit{swap}$ and $\mathit{fetch\&add}$ primitives.
The $n$ low-order bits of the value stored in this object, 
where $n$ is the number of readers, are used to indicate which 
readers have accessed the value stored in the high-order bits.
Each reader is assigned a unique bit, which is set to 1 when the 
reader accesses the value.
Then, when the writer writes a new value it can learn who read its previous value 
by checking the values of the low-order bits it retrieved 
from the value atomically read while writing the new value.
A similar algorithm allows to support multiple auditors, 
but only a single reader,
also using $\mathit{swap}$ and $\mathit{fetch\&add}$.

When there are multiple readers and auditors, 
we use $\mathit{compare\&swap}$, which has consensus number $\infty$,
in addition to $\mathit{fetch\&add}$.

Taken together, 
our results mean that atomic audit cannot be implemented using only 
reads and writes, and stronger primitives (with consensus number 
$> 1$) should be used.  

We then investigate the possibility to extend an atomic register 
with a useful audit operation without relying on strong synchronization
primitives.
This weaker abstraction is called a \emph{regular} audit, and roughly speaking, 
differs from the previous one in not having the audit operations 
linearized together with the write and read operations. 
In particular, a regular audit operation \emph{aop} may not detect a read 
operation that is concurrent with it, 
even though the read has to be linearized before a write that completes 
before the invocation of \emph{aop}.
Our final result is a single-writer multi-reader atomic register with 
multi-auditor \emph{regular} audit,
using only atomic read and write operations, whose consensus number is 1.

\subparagraph*{Organization of the paper.}
After discussing related work in the next section, 
Section~\ref{sec:prelims} describes our model of computation, 
and the definitions of \emph{atomic} and \emph{regular} auditing. 
Sections~\ref{sec:atomic audit consensus} and~\ref{sec:atomic audit imp} contain
the results for an atomic register with atomic audit. 
Section ~\ref{sec:swsrsaConsensus2} elaborates on those results to formalize 
the synchronization power of atomic audit in terms of consensus number.
Section~\ref{sec:reg audit imp} contains the implementation 
of an atomic register with regular audit. 
We conclude in Section~\ref{sec:summary}.




\section{Related Work}

To the best of our knowledge, only two papers~\cite{CB21,SRDS22}  
study auditability of read operations.
Cogo and Bessani~\cite{CB21} were the first to formalize the notion of auditable register. 
Their definition is tailored for auditable register implementations on top of a shared memory model where some base objects can be faulty, i.e., they can omit to record readers or they can record nonexistent read operations. 
They present an algorithm to implement an auditable register,
here read and write operations provide regular semantics 
using $n \geq 4f+1$ atomic read/write shared objects, 
$f$ of which may be faulty. 
Because of their failure model, their high-level register implementation 
relies on information dispersal schemes, where the input of a high-level write 
is split into several pieces each written in a different low-level shared object. 
It implies that a process can read a written value only if it collects enough pieces 
of information, making its read operation detectable. 
Their definition of completeness and accuracy for the auditable register relies 
on the notion of \emph{effectively read}, which they formalize to capture 
the fact that the process executing the high-level read operation could 
have collected enough pieces of information and be able to retrieve the value, 
even if the read operation does not return.

In asynchronous message-passing systems where $f$ processes can be Byzantine, 
Del Pozzo et al.~\cite{SRDS22} study the possibility of implementing 
an atomic auditable register with the accuracy and completeness properties, 
as defined by Cogo and Bessani, with fewer than $4f+1$ processes. 
They prove that without communication between servers, 
auditability requires at least $4f+1$ processes, $f$ of which may be Byzantine. 
They also show that allowing processes to communicate with each other admits 
an auditable atomic register with only $3f+1$ processes, 
providing optimal resilience. 
Their implementation also uses information dispersal scheme to deal with Byzantine processes. 



In contrast, we consider a classical shared memory model where processes can fail by crashing. 
Also, our definition of auditable register is not tailored for a given class of implementations, since it is stated in terms of high-level operations.

Most of the other works on auditing protocols for distributed storage 
focus on data integrity~\cite{WLRY21,DT18,WLL14,DKLT16,TCCJHCL17,LMD15,LHCJXY21},
see a survey in~\cite{KSS19}.
In contrast, our work focuses on auditing \emph{who} has read \emph{which} data.


Frey, Gestin and Raynal~\cite{FreyGR2023} investigate the synchronization power of
\emph{AllowList} and \emph{DenyList}: append-only lists where 
AllowList contains resources that processes can access, 
while DenyList includes resources that processes cannot access. 
They prove the consensus number of AllowList is 1, 
while the consensus number of DenyList is equal to the number of processes 
that can access resources not listed in the DenyList. 
These objects are related to an auditable register, because they 
also record which processes accessed a resource and how many times.
However, the precise nature of this relationship is unclear since some 
variants of an auditable register with atomic audit 
(with a single reader or a single auditor) have conensus number 2, 
a phenomenon that does not happen with AllowList or DenyList.



Finally, we discuss relation of auditability 
to \emph{Accountability}~\cite{peerreview,Civit22Crime, Civit22ABC, DR2022Tenderbake}. 
When faulty processes are malicious, accountability aims to produce proofs of
misbehavior in instances where processes deviate, in an observable way, 
from the prescribed protocol. 
This allows the identification and removal of malicious processes from the system 
as a way to clean the system after a safety violation. 
In contrast, auditability logs the processes actions and let the auditor 
derive conclusions on the processes behavior.

\section{Preliminaries }\label{sec:prelims}

\subsection{Model}

We consider a standard shared-memory model where crash-prone asynchronous processes 
communicate through registers, using a given set of primitive operations. 
The primitive operations (sometimes called just primitives) include 
ordinary $\mathit{read}$ and $\mathit{write}$, 
as well as $\mathit{swap}$, $\mathit{fetch\&add}$ and $\mathit{compare\&swap}$. 

A $\mathit{swap(v)}$ primitive atomically writes $v$ 
to the register and returns its previous value.
A $\mathit{fetch\&add(a)}$ primitive atomically writes the sum of $a$ and the current value of the register into the register and returns its previous value.
A $\mathit{compare\&swap(old,new)}$ primitive is an atomic conditional write: 
the write of $new$ is executed if and only if the value of the register is 
$old$; a boolean value is returned that indicates if the write was successful or not.

The auditable atomic register is an extension of an ordinary 
atomic read/write register~\cite{L86}. 
It is formally defined in the next subsection.
We only consider \emph{single-writer} registers, 
where each register can be written by a single process.

An auditable register implementation specifies the state representation of the register 
and the algorithms processes follow when they perform the read, write and audit operations. 
Each operation has an invocation and a response event.

An execution is a sequence of steps performed by processes as they
follow their algorithms, in each of which a process applies at most a single primitive to the shared memory (possibly in addition to some local computation).  

A \emph{history} $H$ is a sequence of invocation and response events;
no two events occur at the same time. 
An operation is \emph{complete} in history $H$, 
if $H$ contains both the invocation and the matching response for this operation. 
If the matching response is missing, the operation is \textit{pending}.
An operation $op$ \emph{precedes} another operation $op'$ in $H$ 
if the response of $op$ appears before the invocation of $op'$ in $H$;
we also say that $op'$ \emph{follows} $op$.

A history is \emph{sequential} if each operation invocation is immediately 
followed by the matching response, by the same process on the same object. 
For a given history $H$, ${\sf complete}(H)$ is the set of histories 
obtained from $H$ by appending zero or more responses to some pending invocations 
and discarding the remaining pending invocations.

We consider \emph{wait-free} implementations which ensures that  
a non faulty process completes an operation within a finite number of its own steps.






\subsection{Definitions of Auditable Register}
\label{sec:AtomicAuditableRegister}
An auditable register supports three operations:
$R.write(v)$ which assigns value $v$ to the register $R$, $R.read()$ which returns the value of the register, 
and $R.audit()$ which reports the set of all values read in the register and by whom.
Specifically, an audit operation returns a set of pairs $(p,v)$, each corresponding to a read invoked by process $p$ that returned $v$. In the following, we define two specifications for \emph{audit}
operations, exploring different semantics of their interaction 
with concurrent read and write operations. 

\subparagraph*{Atomic Audit.}

Intuitively, atomicity provides the illusion that all the read, write, 
and audit operations appear as if they have been executed sequentially.


\begin{definition}\label{definition:atomic:audit}
A history $H$ is \emph{atomic with atomic audit} if there is 
a history $H^{\prime}$ in ${\sf complete}(H)$
and a sequential history $\pi$ that contains all operations in $H^{\prime}$ such that:
\begin{enumerate}
    \item 
    If operation $\op{1}$ precedes operation $\op{2}$ in $H$, 
    then $\op{1}$ appears before $\op{2}$ in $\pi$.
    \emph{Informally, $\pi$ respects the real-time order of non-overlapping operations.}
    \item 
     Every read in $\pi$ returns the value of the most recent preceding write, 
     if there is one, and the initial value, otherwise.
     \emph{Informally, the history $\pi$ respects the semantics of an 
     atomic read / write register.}

    \item Every audit $\op{}$ in $\pi$ returns a set of pairs $\mathcal{P}$ such that\\
    (Completeness): For each read operation $\mathit{\op{}^\prime}$ by process $p$ that precedes $\op{}$ in $\pi$, $(p,v)\in\mathcal{P}$, where $v$ is the value returned by $\op{}^\prime$.\\
    (Strong Accuracy): For any pair $(p,v) \in \mathcal{P}$, there is a read operation $\op{}^\prime$ by process $p$ that returned $v$, and $\op{}^\prime$ precedes $\op{}$ in $\pi$.
\end{enumerate}
\end{definition}

Roughly speaking, \emph{completeness} formalizes that any read 
of a value from the register must be detected by the audit operation,
while \emph{strong accuracy} ensures that a read is reported by an 
audit operation only if it has occurred.
Note that taken together, completeness and strong accuracy say that 
a pair $(p,v)$ is returned by the audit operation \emph{if and only if}
a read operation by process $p$, returning $v$, 
is linearized in $\pi$ before the audit.
That is, an \emph{atomic audit} operation detects all the read 
operations linearized before it and 
does not detect any read operation linearized after it.  




\subparagraph*{Regular Audit.} 
A \emph{regular audit} operation detects all read operations 
that complete before the audit starts 
and does not detect any read operation that starts after it completes.
An audit operation may detect some subset of the read operations that overlap it.

\begin{definition}\label{definition:regular:audit}
A history $H$ is \emph{atomic with regular audit} if
there is a history $H^{\prime}$ in ${\sf complete}(H)$, 
and a sequential history $\pi$ that contains all read and writes operations in $H^{\prime}$ 
that satisfies the first two conditions of 
Definition~\ref{definition:atomic:audit}, 
and in addition: 
\begin{enumerate}
    \setcounter{enumi}{2}
    \item Every audit $\op{} \in H^\prime$ returns a set of pairs $\mathcal{P}$ such that\\
    (Completeness): For each read operation $\op{}^\prime$ in $H^\prime$
        by process $p$, that completes in $H^\prime$ 
        before the invocation of $\op{}$ in $H^\prime$, 
        $(p,v)\in\mathcal{P}$, where $v$ is the value returned by $\op{}^\prime$.\\
    (Accuracy): For any pair $(p,v) \in \mathcal{P}$, 
        there is a read operation $\op{}^\prime$ $\in$ $H^\prime$ by process $p$ that returned $v$, 
        and the invocation of $\op{}^\prime$ in $H^\prime$ 
        precedes the response of $\op{}$ in $H^\prime$.
\end{enumerate}
\end{definition}

Note that while the condition on atomic audit operations
(Definition~\ref{definition:atomic:audit}) is stated relative 
to the linearization (sequential execution) $\pi$, 
the condition on regular audit is stated relative to 
the completion $H^{\prime}$ of the original history $H$. 
As we shall see, this seemingly-minor change leads to an important 
difference in the synchronization power of audit operations.

\begin{figure}
    \centering
    \includegraphics[scale=0.107]{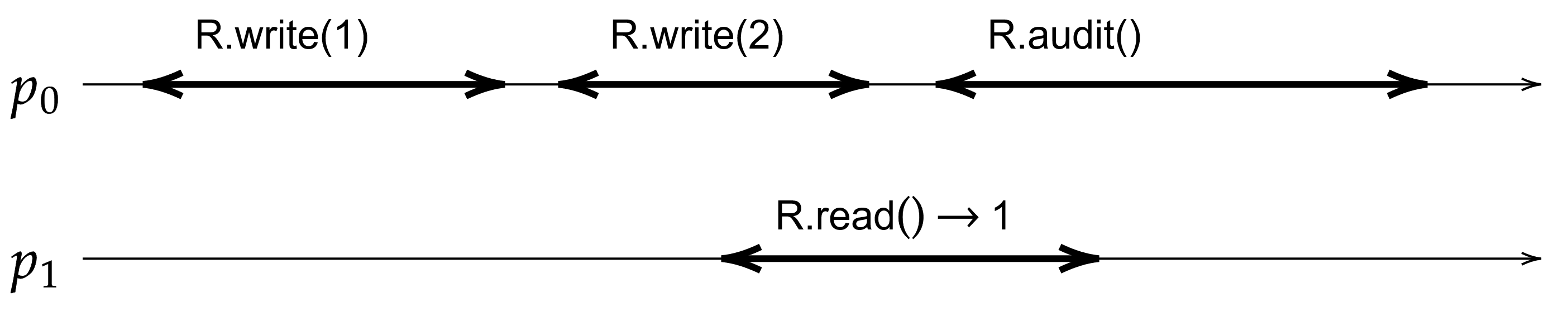}
    \caption{A scenario where a regular audit can return either $\emptyset$ or $(p_1,1)$, 
        while an atomic audit must return $(p_1,1)$.}
    \label{fig:scenarioAtomicRegularRegister}
\end{figure}

Figure~\ref{fig:scenarioAtomicRegularRegister} depicts the difference
between the responses of atomic audit and regular audit.

In the rest of the paper, we consider that only one process can invoke write operations on the register, 
called the \emph{writer}, which is also allowed to invoke audit operations. 
Thus, the \emph{writer} is also an \emph{auditor} of the register. 

When several processes other than the writer are allowed to read the register, 
we call it \emph{multi reader}; otherwise, it is \emph{single reader}. 
Similarly, if several processes other than the writer can audit the register, 
we call it \emph{multi auditor}; otherwise, it is \emph{single auditor}.


\section{Using atomic audit to solve consensus}\label{sec:atomic audit consensus}

In this section, we investigate how atomic audit allows to solve consensus.
An algorithm solving consensus satisfies the following properties:
\begin{description}
    \item[Termination:] A process decides within a finite number of its own steps.
    \label{property:termination}
    \item[Agreement:] All processes decide on the same value.
    \item[Validity:] The decision value has been proposed by some process.
    \label{property:validity}
\end{description} 

\subsection{Single-reader register with single-auditor atomic audit solves two-process consensus}
\label{sec:atomicAuditableConsensusTwoProcess}

Algorithm \ref{alg:async:consensusTwoProcess} solves consensus between two processes 
using two single-writer single-reader (swsr) atomic registers with
a single-auditor atomic audit:
$R_{i}$, for each ${i} \in \{0,1\}$, is a swsr register 
written and audited by process $p_{i}$ and read by $p_{1-i}$.

Each process first writes the value it proposes in its own register. Then it reads the other process's register and audits its own register. 
Finally, it returns its own value or the other process's value, accordingly to the values returned from the read and audit operations. In particular, $p_{i}$ returns its own value (Line~\ref{line:consensusNumberTwo:ReturnReadPerp}) 
if it read the initial value from $R_{1-i}$ 
(Line~\ref{line:consensusNumberTwo:readPerp}). 
In that case, $p_{1-i}$ reads $v_{i}$ from $R_{i}$ (Line~\ref{line:consensusNumberTwo:read}). 
The condition in Line~\ref{line:consensusNumberTwo:readPerp} would not hold, 
and since the audit operation on $R_{1-i}$ detects that $p_{i}$ read $\bot$ from $R_{1-i}$ 
(Line~\ref{line:consensusNumberTwo:audit}), 
$p_{1-i}$ returns the value of $\valueRead$ (Line~\ref{line:consensusNumberTwo:ReturnAuditPerp}),
which is $v_{i}$.
Finally, if $p_{i}$ and $p_{1-i}$ both read the input of the other process 
and they know this fact thanks to the result of the audit operation, 
they apply a deterministic rule to break the tie and choose the same value. 

\begin{algorithm}[tb]

{ 
\begin{algorithmic}[1]
\Statex \textbf{Shared Variables:}
\Statex $R_{i}$, $i\in[0,1]$, swsr atomic register with writer / auditor $p_{i}$ and reader $p_{1-i}$, initially $\bot$ 
\Statex

\Statex \textbf{Local Variables:}
\Statex  $\valueRead$, initially $\perp$ \Comment{value read from $R_{1-i}$}
\Statex  $\valueAudit$, initially $\emptyset$ \Comment{response of audit on $R_{i}$}

\Statex
\START[propose($v_{i}$)]
\Comment{{\bf Pseudo code for process $p_{i}$, $i\in[0,1]$}}
\State{$R_{i}.write(v_{i})$}
\label{line:consensusNumberTwo:write}
\State{$\valueRead \gets$ $R_{1-i}.read()$}
\label{line:consensusNumberTwo:read}
\State{$\valueAudit \gets$ $R_{i}.audit()$}
\label{line:consensusNumberTwo:audit}
%
%
\START[if] {$\valueRead = \perp$}
\label{line:consensusNumberTwo:readPerp}
\State{\textbf{return} $v_{i}$}
\label{line:consensusNumberTwo:ReturnReadPerp}
\EndPhase  
\START[if] {$\valueAudit = (p_{1-i},\perp)$}
\label{line:consensusNumberTwo:auditPerp}
\State{\textbf{return} $\valueRead$}
\label{line:consensusNumberTwo:ReturnAuditPerp}
\EndPhase
\State{\textbf{return} $max(v_{i},\valueRead)$}
\label{line:consensusNumberTwo:ReturnDeter}
\EndPhase
\end{algorithmic}
}

\caption{Two-process consensus using swsr 
atomic registers with single-auditor atomic audit}
\label{alg:async:consensusTwoProcess}
\end{algorithm}


\begin{lemma}\label{lem:async:consensusTwoValidity}
Algorithm~\ref{alg:async:consensusTwoProcess} satisfies validity.
\end{lemma}

\begin{proof}
If $p_{i}$ returns $v_{i}$ 
(Line~\ref{line:consensusNumberTwo:ReturnReadPerp} 
or Line~\ref{line:consensusNumberTwo:ReturnDeter}), 
then validity holds since $v_{i}$ is 
the value $p_{i}$ proposed itself.

Otherwise, $p_{i}$ returns $\valueRead$ 
(Line~\ref{line:consensusNumberTwo:ReturnAuditPerp} 
and Line~\ref{line:consensusNumberTwo:ReturnDeter}). 
Then, $\valueRead$ is the result of a read from $R_{1-i}$ 
(Line~\ref{line:consensusNumberTwo:read}).
We argue that this is the value proposed by $p_{1-i}$. 
Process $p_{i}$ does not satisfy the condition in 
Line~\ref{line:consensusNumberTwo:readPerp}, 
implying that $\valueRead$ is not the initial value of $R_{1-i}$. 
Since the only write operation in $R_{1-i}$ is the input value of $p_{1-i}$ 
(Line~\ref{line:consensusNumberTwo:write}), 
Definition~\ref{definition:atomic:audit}(2) implies validity.
\end{proof}

\begin{lemma}\label{lem:async:consensusTwoAgreement}
Algorithm~\ref{alg:async:consensusTwoProcess} satisfies agreement.
\end{lemma}

\begin{proof}
We consider all possible ways process $p_{i}$ can return a value,
and show that $p_{1-i}$ must return the same value.

\emph{Case 1:} 
$p_{i}$ returns its own value $v_{i}$ 
in Line~\ref{line:consensusNumberTwo:ReturnReadPerp}.
Then, by Line~\ref{line:consensusNumberTwo:readPerp},
$R_{1-i}.read()$ returned $\bot$ to $p_{i}$ 
(Line~\ref{line:consensusNumberTwo:read}). 
Thus, by Definition~\ref{definition:atomic:audit}(1), 
$R_{1-i}.write(v_{1-i})$ by $p_{1-i}$
is linearized after $R_{1-i}.read()$ by $p_{i}$.
This implies that $R_{i}.write(v_{i})$ is linearized before $R_{i}.read()$,
which returns $v_{i}$ to $p_{1-i}$, 
by Definition~\ref{definition:atomic:audit}(2). 
Finally, by Definition~\ref{definition:atomic:audit}(3),
$R_{1-i}.audit()$ returns $\{(p_{i},\bot)\}$ to $p_{1-i}$.
Therefore, the condition in Line~\ref{line:consensusNumberTwo:auditPerp}
holds for $p_{1-i}$ and it returns $\valueRead$ that contains $v_{i}$ 
(Line~\ref{line:consensusNumberTwo:ReturnAuditPerp}).

\emph{Case 2:} 
$p_{i}$ returns $\valueRead$ at Line~\ref{line:consensusNumberTwo:ReturnAuditPerp}, 
which holds $v_{1-i}$, the value proposed by $p_{1-i}$. 
In this case, the condition at Line~\ref{line:consensusNumberTwo:auditPerp} 
holds for $p_{i}$.
By Definition~\ref{definition:atomic:audit}(3), 
$R_{i}.read()$ by $p_{1-i}$ returns $\perp$ (Line~\ref{line:consensusNumberTwo:read}),
implying that $p_{1-i}$ returns $v_{1-i}$ 
(by Line~\ref{line:consensusNumberTwo:readPerp} 
and Line~\ref{line:consensusNumberTwo:ReturnReadPerp}).

\emph{Case 3:} 
$p_{i}$ returns at Line~\ref{line:consensusNumberTwo:ReturnDeter}.
This means that $p_{i}$ has read $v_{1-i}$ from $R_{1-i}$ 
(Line~\ref{line:consensusNumberTwo:read}) 
and its audit response is not $\{(p_{1-i} ,\perp)\}$ 
(Line~\ref{line:consensusNumberTwo:audit}). 
This means that the audit response is either $\emptyset$ meaning 
that $p_{1-i}$ has not yet read $R_i$ or $(p_{1-i}, v_i)$.  
By Definition~\ref{definition:atomic:audit}(3),  
$R_{i}.read()$ by $p_{1-i}$ is linearized after 
$R_{i}.audit()$ by $p_{i}$. 
Thus, $R_{i}.write(v_{i})$ by $p_{i}$ is linearized before 
$R_{i}.read()$ by $p_{1-i}$, which therefore, 
returns $v_{i}$ to $p_{1-i}$ (Line~\ref{line:consensusNumberTwo:read}).
Since $R_{1-i}.read()$ by $p_{i}$ did not return $\bot$, 
Definition~\ref{definition:atomic:audit}(3) implies that $\{(p_{i},\bot)\}$
is not the response of $R_{i}.audit()$ by $p_{i}$.
It follows that $p_{i}$ and $p_{1-i}$ return the same value which is the maximum 
between $v_{i}$ and $v_{1-i}$.
\end{proof}


The algorithm satisfies validity (Lemma~\ref{lem:async:consensusTwoValidity})
and agreement (Lemma~\ref{lem:async:consensusTwoAgreement}). 
Furthermore, \textbf{propose} performs a constant number of primitive operations, 
so Algorithm~\ref{alg:async:consensusTwoProcess} is clearly wait-free, 
therefore respects the termination property. 
This implies:

\begin{theorem}
\label{theorem:async:consensusTwoProcesses}
Algorithm~\ref{alg:async:consensusTwoProcess} solves consensus for two processes.
\end{theorem}

\subsection{Multi-reader register with multi-auditor atomic audit solves $n$-process consensus}
\label{sec:atomicAuditableConsensusNProcesses}

We now generalize Algorithm~\ref{alg:async:consensusTwoProcess}
to solve consensus among $n$ processes 
using single-writer multi-reader (swmr) atomic registers with 
\emph{multi-auditor} atomic audit. 
Like the algorithm for two-process consensus,
processes leverage the audit to reconstruct at which point of the execution 
the other processes are, and base their decision on it.  

Algorithm~\ref{alg:async:consensusInfinite} uses 
$n$ swmr atomic registers with multi-auditor atomic audit 
$R_0$, \ldots, $R_{n-1}$, all initially $\bot$.
Process $p_{i}$ is the single writer of $R_{i}$,  
and all processes can read and audit $R_{i}$.

Each process $p_{i}$ proposes its input, 
by writing it in $R_{i}$. 
Then, $p_{i}$ reads and audits all the other registers. 
A simple situation is when one process, say $p_{i}$,
writes and reads before all other processes,  
the audit detects that $p_{i}$ read $v_{i}$ in $R_{i}$ and $\bot$ in all other registers. 
This implies that all later processes will read $p_{i}$'s value in $R_{i}$ and, 
thanks to the audit, detect that $p_{i}$ is not aware of the other processes' propositions. 
In this case, $v_{i}$ is the only value known to all processes, 
so it is safe to decide on $v_{i}$.

In general, we can consider the set $P$ of processes that write 
before any process reads. 
No process reads $\bot$ from the registers of processes in $P$, 
and this can be detected by auditing these registers.
This means that all processes consider the input values of processes 
in $P$ as safe to decide upon, and agreement can be reached 
by deterministically picking one of these values, 
e.g., the maximal one.

Each process keeps the following local data structures:
$\Values[\,]$ is an array of length $n$ to hold the values read from $R_0$, \ldots, $R_{n-1}$, initially $\perp$;
$\SafeValues$ is a set that stores the proposed values that no process missed, initially $\emptyset$; and
$\AuditResponse$ holds the results of audits on $R_0$, \ldots, $R_{n-1}$, initially $\emptyset$.

When proposing a value $v_{i}$, each process $p_{i}$ first writes $v_{i}$ 
in $R_{i}$ (Line~\ref{line:consensusNumberInfinite:write}). 
Next, $p_{i}$ reads $R_0$, \ldots, $R_{n-1}$ and stores the responses 
in $\Values[\,]$ (Line~\ref{line:consensusNumberInfinite:read}).
Finally, $p_{i}$ audits $R_0$, \ldots, $R_{n-1}$ and stores  
the returned pairs in a set $\AuditResponse$ 
(Line~\ref{line:consensusNumberInfinite:audit}). 
For each $R_{j}$, when a value is added to $\AuditResponse$, 
$p_{i}$ checks if there is a process that read $\bot$ from $R_{j}$ 
(Line~\ref{line:consensusNumberInfinite:conditionNoPerp}). 
If this is not the case, then the value in $R_{j}$ is considered safe, 
and is added to $\SafeValues$ 
(Line~\ref{line:consensusNumberInfinite:safeValue}).
Finally, it returns the maximum value in $\SafeValues$ 
(Line~\ref{line:consensusNumberInfinite:return}).
(We assume that the input values are from a totally-ordered set.)

\begin{algorithm}[tb]

{ 
\begin{algorithmic}[1]
\Statex \textbf{Shared Variables:}
\Statex $R_i$, $i\in[0,n-1]$, 
swmr atomic registers with multi-auditor atomic audit;
$R_{i}$ is written by process $p_{i}$, initially $\bot$

\Statex
\Statex \textbf{Local Variables:}
\Comment{{\bf Pseudo code for process $p_{i}$, $i\in[0,n-1]$}}
\Statex $\Values[\,]$  an array of length $n$, initially $\perp$ 
\Statex $\SafeValues$ a set, initially $\emptyset$ 
\Statex $\AuditResponse$ a set, initially $\emptyset$ 

\START[propose($v_{i}$)]
\State{$R_{i}.write(v_{i})$}
\label{line:consensusNumberInfinite:write}
\START[for] $0\leq {j} < n$
    \State{$\Values[{j}] \gets$ $R_{j}.read()$}
    \label{line:consensusNumberInfinite:read}
\EndPhase

\START[for] $0\leq {j} < n$
    \State{$\AuditResponse \gets R_{j}.audit()$}
    \label{line:consensusNumberInfinite:audit}
    \START[if]{ $\not\exists (*,\perp) \in \AuditResponse$} 
    \label{line:consensusNumberInfinite:conditionNoPerp}
    \Comment{no process read $\perp$ from $R_{j}$}
        \State{$\SafeValues.add(\Values[{j}]$)}
    \label{line:consensusNumberInfinite:safeValue}
    \EndPhase
\EndPhase
\State{\textbf{return} max($\SafeValues$)}
\label{line:consensusNumberInfinite:return}
\EndPhase

\end{algorithmic}
}

\caption{$n$-process consensus using swmr 
atomic registers with multi-auditor atomic audit}
\label{alg:async:consensusInfinite}
\end{algorithm}

\begin{lemma}\label{lem:async:consensusInfinite2}
A process $p_i$ adds a value $v$ to $\SafeValues$, 
only if $v$ was proposed by some process.
\end{lemma}

\begin{proof}
Process $p_i$ adds values it reads from $R_0$, \ldots, $R_{n-1}$, 
the registers of other processes, to $\SafeValues$ in 
Line~\ref{line:consensusNumberInfinite:safeValue}.
The value read from a register $R_j$, 
in Line~\ref{line:consensusNumberInfinite:read}, 
is either $\perp$ (the initial value of the register) 
or the value proposed by $p_j$, written to $R_j$ 
in Line~\ref{line:consensusNumberInfinite:write}. 

We next argue that $p_i$ does not add $\bot$ to $\SafeValues$.
If $p_i$ read $\bot$ from some $R_j$,
then since its audit on $R_j$ follows its read from $R_j$,
the Completeness of its audit operation
(Definition~\ref{definition:atomic:audit}(3))
implies that $(p_i,\bot)$ is contained in the response of 
$R_j.{\swmrmaIaudit}()$. 
By the condition of Line~\ref{line:consensusNumberInfinite:conditionNoPerp},
$\bot$ is not added to $\SafeValues$.
\end{proof}

\begin{lemma}\label{lem:async:consensusInfiniteValidity}
Algorithm~\ref{alg:async:consensusInfinite} satisfies validity.
\end{lemma}

\begin{proof}
A process decides on a value in $\SafeValues$ 
(Line~\ref{line:consensusNumberInfinite:return}),
and by Lemma~\ref{lem:async:consensusInfinite2},
this set contains only values proposed by some process. 
We complete the proof by showing that $\SafeValues$ is not empty.

Let $p_k$ be the first process to complete its {\swmrmaIwrite} 
of $v_k$ to $R_k$. 
Since all processes read the registers of the other processes
after finishing the {\swmrmaIwrite}, 
it follows that all the processes read $v_k \neq \bot$ from $R_k$.
By Strong Accuracy (Definition~\ref{definition:atomic:audit}(3)), 
the {\swmrmaIaudit} of $R_k$ does not contain a pair 
$(p_j,\perp)$, for any $p_j$. 
Therefore, 
the condition in Line~\ref{line:consensusNumberInfinite:conditionNoPerp}
holds and $p_i$ adds $v_k$ to $\SafeValues$ 
in Line~\ref{line:consensusNumberInfinite:safeValue}, as needed.
\end{proof}

\begin{lemma}\label{lem:async:consensusInfiniteAgreement}
Algorithm~\ref{alg:async:consensusInfinite} satisfies agreement.
\end{lemma}

\begin{proof}
We prove that all processes have the same set $\SafeValues$ when 
deciding, which immediately implies agreement. 
Suppose that process $p_i$ is the first to add a value $v_k$ 
to its $\SafeValues$ set. 
This means that $p_i$ reads $v_k$ from register $R_k$ 
(Line~\ref{line:consensusNumberInfinite:read}).

Let $aop_i$ be the audit by $p_i$ on $R_k$ 
(Line~\ref{line:consensusNumberInfinite:conditionNoPerp}).
Since $p_i$ adds $v_k$ to $\SafeValues$, 
it follows that no pair $(p_j,\bot)$, for some process $p_j$, 
is included in the response to $aop_i$.
This implies that all read operations from $R_k$ that are linearized
before $aop_i$ do not return $\bot$.

Consider a read operation by process $p_j$ from $R_k$ that is 
linearized after $aop_i$. 
This follows the read of processes $p_i$ from $R_k$, 
which returns $v_k \neq \bot$,
and hence, this read will also read $v_k \neq \bot$.
(Since only $p_k$ writes to $R_k$, once, 
changing its value from $\bot$ to $v_k$.)

Thus, no read from $R_k$ returns $\bot$.
This means that any process $p_i'$ consider $v_k \neq \bot$ 
in Line~\ref{line:consensusNumberInfinite:conditionNoPerp}.
Moreover, by the Strong Accuracy property of the audit operation 
(Definition~\ref{definition:atomic:audit}(3)), 
it follows that no pair $(p_j,\bot)$ is contained in the result 
of the audit operation by $p_i'$ on $R_k$.
This implies that $v_k$ is in $\SafeValues$ of $p_i'$.

This implies that the $\SafeValues$ sets of all processes 
are identical, and they all decide on the same value. 
\end{proof}

Therefore, the algorithm
satisfies validity (Lemma~\ref{lem:async:consensusInfiniteValidity})
and agreement (Lemma~\ref{lem:async:consensusInfiniteAgreement}). 
Furthermore, all the loops in \textbf{propose} are iterated at most $n$ times.
Since the operations invoked in \textbf{propose} are wait-free, 
we get that Algorithm~\ref{alg:async:consensusInfinite} is wait-free.
This implies:


\begin{theorem}\label{theorem:async:consensusAnyProcesses}
Algorithm~\ref{alg:async:consensusInfinite} solves consensus for $n$ processes.
\end{theorem}


\section{Atomic audit implementations}\label{sec:atomic audit imp}

We now turn to present several implementations of an atomic single-writer 
register with atomic audit. 
The results of the previous section indicate which synchronization primitives 
must be used in the implementations.
Since two-process consensus can be solved with a single auditor and single reader, 
we cannot avoid synchronization primitives with consensus number at least two;
we use swap and fetch\&add 
(Sections~\ref{sec:Implementationswsrsa},~\ref{sec:Implementationswmrsa} 
and~\ref{sec:Implementationswsrma}).
When there are multiple auditors and multiple readers,
we use a universal synchronization primitive, compare\&swap (Section~\ref{sec:casAlgorithm}),
in addition to fetch\&add.

\subsection{Implementing single-reader atomic register with single-auditor atomic audit using swap}
\label{sec:Implementationswsrsa}

We implement a swsr atomic register with single-auditor atomic audit
using a swap primitive.
We use a shared register $R$ that holds the last written value, 
if the last operation was a write, 
or a special value ($\perp$) if the last operation was a read;
the audit operations do not affect the value of $R$.
In a write($v)$ operation, the (single) writer applies $\mathit{swap}$ to $R$, 
atomically writing $v$ into $R$ and retrieving the overwritten value 
to check if the reader read the previously written value. 
In the latter case, the swap returns a special value $\bot$. 

The pseudo-code appears in Algorithm~\ref{alg:AtomicAuditableRegister:swsrsa}. 
The reader keeps the following local data structures:
$\temp$ that holds the value read from $R$, initially $\perp$; and
$\readResult$ that holds the value returned by the last read operation, initially $\perp$.
The writer and auditor keeps the following local data structures:
$\currVal$ that hold the last value written in $R$, initially $v_0$;
$\prevVal$ that hold the previous value written in $R$, initially $\perp$; and
$\auditResult$ a set that stores the pairs (process,value) of the detected read operation, initially $\emptyset$.

In a read, the reader atomically reads the last value written 
into $R$ and swaps it with $\bot$ to notify the writer 
that it read the last value written.
If the response is not $\bot$, then this is the response of the read,
which the reader stores in $\readResult$ for future read operations,
before returning. 
Otherwise, no write has occurred since its previous read, 
so the read returns the value in $\readResult$
(without changing its value). 

In a write, the (single) writer stores in $\prevVal$ the value of 
the previous write, from $\currVal$ (Line~\ref{line:swsrsa:writePrevVal}),
and stores in $\currVal$ the value $v$ it is going to write
(Line~\ref{line:swsrsa:writeCurVal}).
In this way, if the next write operation detects that the reader has read 
the previous value written, the writer knows what this value is. 
Then, the writer swaps $\currVal$ into $R$:
atomically writing it into $R$ and retrieving the overwritten value.
If the writer gets $\bot$ from the swap, then the reader has 
read the last value it wrote (stored in $\prevVal$), and it adds the pair 
(reader, $\prevVal$) to $\auditResult$ (Line~\ref{line:swsrsa:writeAudit}).

In an audit, the auditor (who is also the writer) returns all 
the (process,value) pairs collected during the previous write operations. 
By reading $R$, the auditor checks whether the reader read the value of  
the last write operation, in which case $R$ is $\bot$.
In this case, it adds the pair (reader, $\currVal$) to $\auditResult$. 
Finally, the audit returns $\auditResult$.

The algorithm uses a shared variable $R$ that holds either a value or a special character.  

\begin{algorithm}[tb]

{ 
\begin{algorithmic}[1]
\Statex \textbf{Shared Variables:}
\Statex $R$, accessed with \textit{read} and \textit{swap}, initially $v_{0}$

\Statex
\Statex \textbf{Local Variables:}
\Comment{{\bf Pseudo code for reader $p_r$}}
\Statex $\temp$, initially $\perp$ \Comment{result of the swap}
\Statex $\readResult$, initially $\perp$ \Comment{value returned by the read}

\Statex
\START[Read()]
    \State {$\temp \gets$ $R.swap(\perp)$} \label{line:swsrsa:readSwap}
    \START[if] $\temp  \neq \perp$
    \label{line:swsrsa:conditionRead}
        \State {$\readResult \gets \temp $}
        \label{line:swsrsa:readResultUpdate}
    \EndPhase
    \State \textbf{return} $\readResult$
    \label{line:swsrsa:readReturn}
\EndPhase

\Statex
\Statex \textbf{Local Variables:}
\Comment{{\bf Pseudo code for writer and auditor $p_w$}}
\Statex $\currVal$, initially  $v_0$ \Comment{last value written}
\Statex $\prevVal$, initially  $\perp$ \Comment{previous value written} 
\Statex $\auditResult$, initially $\emptyset$ 
        \Comment{set of tuples $(p,v)$, with $p$ the reader and $v$ a value} 

\Statex
\START[Write($v$)]
    \State {$\prevVal \gets \currVal$}
    \label{line:swsrsa:writePrevVal} 
    \State {$\currVal \gets v$}
    \label{line:swsrsa:writeCurVal} 
    \START[if] $R.swap(v)=\perp$
    \label{line:swsrsa:writeSwap} 
        \State{$\auditResult.add \gets (p_r,\prevVal)$}
    \label{line:swsrsa:writeAudit} 
    \EndPhase
    \State{\textbf{return}}
\EndPhase

\Statex
\START[Audit()]
    \START[if] $R.read()=\perp$ 
    \label{line:swsrsa:auditCondition} 
        \State{$\auditResult.add (p_r,\currVal)$}
        \label{line:swsrsa:auditAddAudit} 
    \EndPhase
    \State \textbf{return} $\auditResult$
    \label{line:swsrsa:auditReturn}
\EndPhase

\end{algorithmic}
}

\caption{Implementation of a single-reader atomic register 
with single-auditor atomic audit using swap}
\label{alg:AtomicAuditableRegister:swsrsa}
\end{algorithm}

\subparagraph*{Proof of Correctness:} 
We assume that the values written to the register are unique.

Fix a history $H$. 
It has at most two pending operations: 
one, either an audit or a write, by process $p_w$,
and another by process $p_r$, which must be read.
We never complete a pending audit.
We complete a pending read in $H$ if and only if some audit contains $(p_r,v)$ 
in its response and no preceding read in $H$ (which must be complete) returns $v$.
We complete a pending write in $H$ if and only if some read (including those 
completed) returns the corresponding value.

A pending operation that is completed has accessed $R$ with a swap:
A read is completed if it is the only read that returns a value detected by an 
audit, thus, the read has executed the swap in Line~\ref{line:swsrsa:readSwap}.
A write is completed if some read has read its value, namely, 
the write has executed the swap in Line~\ref{line:swsrsa:writeSwap}. 
 
We totally order all the completed operations by the order they apply 
their unique swap on $R$.
Call this total order $\pi$ and note that it respects the real-time order 
of the high-level operations on the register, since the swaps are inside
the operations' intervals. 

\begin{lemma}\label{lem:swsrsa2}
Every read in $\pi$ returns the value of the most recent preceding write, 
if there is one, and the initial value, otherwise. 
\end{lemma}
    
\begin{proof}
Consider a read $\op{r}$ that returns a value $v$,
and let $\op{r}^{\prime}$ be the first read that returns this value.
Since $\readResult$ is updated only if the value returned by the swap 
in Line~\ref{line:swsrsa:readSwap} is not $\bot$,
then the swap of $\op{r}^{\prime}$ returns $v$.
Thus, there is a preceding swap that sets $R$ to $v$, 
and it must be by some write $\op{w}$ of value $v$.
Since reads and writes are linearized by the order of their swaps,
$\op{w}$ precedes $\op{r}^{\prime}$, and therefore, 
also $\op{r}$ in $\pi$.

We next argue that no other write is linearized between $\op{w}$ and 
$\op{r}$ in $\pi$. 
Assume otherwise, and let $\op{w}^{\prime}$ be the last write that is linearized 
before $\op{r}$ in $\pi$. 

If the swap of $\op{r}$ in Line~\ref{line:swsrsa:readSwap} returns a value 
different from $\bot$, then this value was written by $\op{w}^{\prime}$ 
because this is the last preceding swap that writes a non-$\bot$ value 
before the swap by $\op{r}$.
This contradicts the assumption that $\op{r}$ returns the value written 
by $\op{w}$.

If the swap of $\op{r}$ in Line~\ref{line:swsrsa:readSwap} returns $\bot$,
this means that an earlier read that executed Line~\ref{line:swsrsa:readSwap} 
after $\op{w}^{\prime}$ executed its swap. 
The first such read swaps from $R$ the value written by $\op{w}^{\prime}$ with $\bot$.  
By Line~\ref{line:swsrsa:readResultUpdate}, the value of $\readResult$ 
is not $v$ when $\op{r}$ returns in Line~\ref{line:swsrsa:readReturn},
which is a contradiction.
\end{proof}

\begin{lemma}\label{lem:swsrsa:completeness}
The set of pairs $\mathcal{P}$ returned by an audit in $\pi$ 
satisfies the Completeness property. 
\end{lemma}

\begin{proof}
Consider an audit $\op{a}$ that returns a set $\mathcal{P}$,
and let $\op{r}$ be a read returning $v$ that precedes $\op{a}$ in $\pi$.
By Lemma~\ref{lem:swsrsa2}, every read returns the value of the 
most recent preceding write in $\pi$. 
Let $\op{r}^{\prime}$ be the first read that returns $v$.
Then $\op{r}^{\prime}$ sets $R$ to $\perp$, and 
the value in $\currVal$ is $v$.

If there is no write between $\op{r}$ and $\op{a}$, 
the audit reads $\bot$ from $R$ (Line~\ref{line:swsrsa:auditCondition}),
while $\currVal$ is still $v$ in Line~\ref{line:swsrsa:auditAddAudit},
implying that the audit adds $(p_r,v)$ to $\mathcal{P}$. 
Otherwise, there is a write between $\op{r}$ and $\op{a}$.
Let $\op{w}$ be the first such write, and notice that $\op{w}$ completes,
since there is a following audit (by the same process). 
Moreover, since it is the first write after $\op{r}$, 
the value of $R$ is $\perp$ when $p_w$ executes Line~\ref{line:swsrsa:writeSwap} 
and $\currVal$ is $v$ immediately before it executes Line~\ref{line:swsrsa:writePrevVal}. 
Thus, the pair $(p_r,v)$ is added to $\auditResult$.
\end{proof}

\begin{lemma}\label{lem:swsrsa:strongAccuracy}
The set of pairs $\mathcal{P}$ returned by an audit in $\pi$ 
satisfies the Strong Accuracy property. 
\end{lemma}

\begin{proof}
Consider an audit $\op{a}$ that returns a set $\mathcal{P}$,
and let $(p_r,v)$ be a pair in $\mathcal{P}$. 
The first operation $\op{}$ that adds $(p_r,v)$ to $\mathcal{P}$ is either 
$\op{a}$ itself or a write / audit that precedes $\op{a}$ in $\pi$. 
This is because the variable $\readResult$ holding set $\mathcal{P}$ is 
updated immediately after the swap by $p_w$ in the corresponding operation. 

If $(p_r,v)$ is added to $\mathcal{P}$ by an audit $\op{}$, 
then $\currVal$ is $v$ when this happens. 
Since the condition in Line~\ref{line:swsrsa:auditCondition} holds, 
there is a reads that swaps $\perp$ into $R$ after $v$ was written to $R$. 
This read is between the write of $v$ and $\op{}$ and by Lemma~\ref{lem:swsrsa2},
it returns $v$, proving the lemma.

If $(p_r,v)$ is added to $\mathcal{P}$ by a write $\op{}$,
then by Line~\ref{line:swsrsa:writePrevVal} and Line~\ref{line:swsrsa:writeAudit}, 
$v$ is the value written by the write that immediately precedes 
$\op{}$ in $\pi$.
Then $v$ is the value returned by the read that swaps $v$ with $\perp$,
which allows the condition in Line~\ref{line:swsrsa:writeSwap} to hold. 
This read precedes $\op{}$ and therefore, it also precedes $\op{a}$.
\end{proof}

Lemma~\ref{lem:swsrsa2}, Lemma~\ref{lem:swsrsa:completeness} 
and Lemma~\ref{lem:swsrsa:strongAccuracy} imply:

\begin{theorem} \label{theorem:swsrsa:correctness}
Algorithm~\ref{alg:AtomicAuditableRegister:swsrsa} implements
a single-writer single-reader atomic register with single-auditor atomic audit.
\end{theorem}


\subsection{Implementing multi-reader atomic register with single-auditor atomic audit using swap and fetch\&add}
\label{sec:Implementationswmrsa}

The algorithm for a \emph{multi-reader} atomic register 
with single-auditor atomic audit follows a similar idea as the 
algorithm for a \emph{single} reader in the previous section,
by having each reader leave a trace of each of its reads.
However, there is an additional difficulty of allowing the writer 
to atomically retrieve the traces of all readers 
when writing a new value or doing an audit.

We address this difficulty by using fetch\&add, 
in addition to swap. 
A $\mathit{fetch\&add}$ allows to accurately change the value of a shared variable 
$R$ so that its binary representation captures multiple pieces of information: 
The high-order bits hold the value written by the writer,
while the $n$ low-order bits indicate whether the readers have read the value.
Specifically, the bit in position $i$, denoted $\bit_i$, is associated with 
reader $p_i$, $0 \leq i < n$, and holds either $0$ or $1$. 
$\bit_i$ is set to $1$ by $p_i$ to indicate that it has read 
the value stored in the high-order bits of $R$; it is $0$, otherwise.
We use two functions to extract information from $R$. 
If $R$ holds $temp$, 
then $\mathsf{GetValue}(temp)$ retrieves the value stored in the high-order bits of $temp$ 
and $\mathsf{GetsBits}(temp)$ retrieves an array of $n$ low-order bits of $temp$.

\begin{algorithm}

{ 
\begin{algorithmic}[1]

\Statex \textbf{Shared Variables:}
\Statex  $R$ 
accessed with $\mathit{read}$, $\mathit{swap}$ and $\mathit{fetch\&add}$ primitives, initially $v_0*2^n$

\Statex
\Statex \textbf{Local Variables:}
\Comment{{\bf Pseudo code for reader $p_i,\, i\in[0,n-1]$}}
\Statex $\tempMr $, initially $\perp$ \Comment{content of the register}
\Statex $read\_result$, initially $\bot$ \Comment{last value read}
\Statex

\START[Read()] 
    \State{$\tempMr \gets R.read()$}
        \label{line:swmrsa:read}
    \START[if]{ $(GetBits(\tempMr)[i] = 0)$} 
     \label{line:swmrsa:readCondition}
    \State{$read\_result \gets GetValue(R.fetch\&add(2^i))$}
        \label{line:swmrsa:readFetch}
    \EndPhase
    \State{\textbf{return} $read\_result$}
        \label{line:swmrsa:readReturn}
\EndPhase


   


   

\Statex
\Statex \textbf{Local Variables:}
\Comment{{\bf Pseudo code for writer and auditor $p_w$}}
\Statex $\auditResult$, initially $\emptyset$ 
\Comment{set of tuples $(p,v)$, with $p$ the reader and $v$ a value}
\Statex $\currVal$, initially $v_0$ \Comment{last value written}
\Statex $\prevVal$, initially $\perp$ \Comment{previous value written} 
\Statex $\tempWriteMr$ with initial value $\perp$ \Comment{content of the register} 

\Statex
\START[Write($v$)]
    \State {$\prevVal \gets \currVal$}
    \label{line:swmrsa:writePrevVal} 
    \State {$\currVal \gets v$}
    \label{line:swmrsa:writeCurVal} 
    \label{line:swmrsa:writeHighOrder}
    \State {$\tempWriteMr \gets R.swap(v,0^{\Nreader})$}
    \Comment{write $v$ in the high order bits}
 \label{line:swmrsa:writeSwap}
    \START[for] $0 \leq j < \Nreader$
\label{line:swmrsa:writemodulo}
    \START[if] $(GetBits(\tempWriteMr)[j]=1)$
    \label{line:swmrsa:writecheckone}
    \Comment{check if $p_j$ read the previous value}
            \State{$\auditResult.add(p_j,\prevVal)$}
            \label{line:swmrsa:writeAudit}
        \EndPhase
    \EndPhase
    \State{\textbf{return}}
    
\EndPhase

\Statex

\START[Audit()]
    \State {$\tempWriteMr \gets R.read()$}
    \label{line:swmrsa:auditRead}

    \START[for] $0 \leq j < \Nreader$
        \START[if] $(GetBits(\tempWriteMr)[j] = 1)$
         \Comment{check if $p_j$ read the last value}
         \label{line:swmrsa:auditCondition}
            \State{$\auditResult.add(p_j,\currVal)$}  
            \label{line:swmrsa:auditAddAudit}
        \EndPhase
    \EndPhase
    \State{\textbf{return} $\auditResult$}
    \label{line:swmrsa:auditReturn}
\EndPhase

\end{algorithmic}
}

\caption{Implementation of multi-reader atomic register 
with single-auditor atomic audit using $\mathit{fetch\&add}$ and $\mathit{swap}$,
for $\Nreader$ readers}
\label{alg:AtomicAuditableRegister:swmrsa}
\end{algorithm}

In more detail (see Algorithm~\ref{alg:AtomicAuditableRegister:swmrsa}), 
when a reader $p_i$ reads a value written in $R$, 
it sets $\bit_i$ to $1$ by adding $2^i$ to the value stored in $R$. 
Since a reader can read the same value several times, 
$p_i$ checks that $\bit_i$ is not already set to $1$, 
before adding $2^i$ to $R$ (Line~~\ref{line:swmrsa:readFetch}). 
This ensures that $p_i$ changes only its bit. 

When writing a new value $v$, 
the writer 
swaps the value $v$ and resets the $n$ low-order bits to 0
into $R$ and 
obtains the previous value of $R$, into a local variable \emph{$\tempMr$}.
Then for each reader $p_i$, the writer retrieves $\bit_i$ from \emph{$\tempMr$}
(Lines~\ref{line:swmrsa:writemodulo} and~\ref{line:swmrsa:writecheckone}). 
If $\bit_i$ is equal to $1$, the writer knows that reader $p_i$ has read 
the previous value and the pair $(p_i, \prevVal)$ 
is added to the set to be returned by an audit, called $\auditResult$.
$\auditResult$ is a local variable, which can be accessed both by 
the writer and the auditor because they are the same process. 

In a similar manner, an audit operation also reads $R$ to detect 
high-level Read operations that may have read the last value written 
but that has not been detected by the last write operation.
Since $\auditResult$ is a set, 
the pair will not be added if it was already in the set. 
(An efficient implementation of a sequential set can be used for this local variable.)

Note that the algorithm uses a shared variable $R$ that holds a value and $n$ bits.

\subparagraph*{Proof of Correctness:}
We assume that the values written to the register are unique.

Fix a history $H$. Note that there are at most $\Nreader+1$ pending operations 
in $H$: one (either an audit or a write) by the writer, 
and possibly one read operation for each reader.
We never complete a pending audit. 
We complete a pending read invoked by process $p_i$ in $H$ if and only if 
some audit contains $(p_i,v)$ in its response and 
no earlier read in $H$ (which must be complete) returns $v$ to $p_i$.
We complete a pending write in $H$ if and only if some read in $H$ returns the corresponding value.
Note that if a pending operation is completed, then it applied a primitive to $R$:
a write is completed if some read has read its value, namely, 
the write has executed the swap in Line~\ref{line:swmrsa:writeSwap}; 
a read is completed if it is the only read that returns a value detected by an 
audit, thus, the read has executed the $\mathit{fetch\&add}$ in Line~\ref{line:swmrsa:readFetch}.
 
We totally order all the completed operations according to the 
order they apply their last primitive ($\mathit{swap}$, $\mathit{read}$ or $\mathit{fetch\&add}$)  
to $R$. 
A write or an audit apply only one primitive. 
For a read, the last primitive is the $\mathit{fetch\&add}$, 
if this is the first time that the process reads a given value, 
and otherwise, it is the read.
Let $\pi$ denote this total order, 
and note that it respects the real-time order of 
the high-level operations on the auditable register 
because each such step is in the execution interval of the corresponding operation. 

\begin{lemma}\label{lemma:setRead}
For each $i \in\{0,\ldots n-1\}$, $\bit_i$ is set to $0$ each time a write operation executes Line~\ref{line:swmrsa:writeSwap} and it is set to $1$ if and only if a read operation by $p_i$ executes Line~\ref{line:swmrsa:readFetch}.
\end{lemma}
\begin{proof}
When a write operation executes Line~\ref{line:swmrsa:writeSwap}, it writes a value $v*2^{n}$ into $R$ for some integer $v$. Thus the $n$ less significant bits of the binary representation of the value stored in $R$ are set to $0$.
For each $i \in\{0,\ldots n-1\}$, process $p_i$ adds the amount $2^{i}$ to the value stored in $R$ if and only if the $i$-th bit of its binary representation is equal to $0$. Thus, $p_i$ sets the $i$-th bit to $1$ and does not change any other bit of $R$.
\end{proof}

\begin{lemma}\label{lem:swmrsa2}
Every read in $\pi$ returns the value of the most recent preceding write in $\pi$, 
if there is one, and the initial value, otherwise. 
\end{lemma}
    
\begin{proof}
Consider a read $\op{r}$ by a process $p$ that returns a value $v$. 

If the condition in Line~\ref{line:swmrsa:readCondition} holds, 
then the result of $\op{r}$ is the value extracted from the value read 
by applying the fetch\&add primitive on $R$ 
(the whole content of $R$ without the bits dedicated for auditing) 
in Line~\ref{line:swmrsa:readFetch}. 
Thus, there is a preceding swap primitive that set the high-order bits 
of $R$ to $v$, and it must be by some write operation $\op{w}$.
Since reads and writes are linearized according to the order 
they apply their last primitive to $R$,
$\op{w}$ precedes $\op{r}$ in $\pi$. For the same reason, 
$\op{w}$ should be the most recent preceding write in $\pi$. 
Otherwise, $\op{r}$ would have returned a different value.

Otherwise, the condition at Line~\ref{line:swmrsa:readCondition} does not hold, 
then the value returned by $\op{r}$ has been computed by the most recent read operation that precedes $\op{r}$ in the program order of $p$ for which the condition evaluates to $true$, denoted $\op{r'}$. Also, by Lemma~\ref{lemma:setRead} and since $\bit_i\neq 0$, no write changed the values of $R$ after $\op{r'}$ and before $\op{r}$. Thus, the claim follows from the previous case.
\end{proof}

\begin{lemma}\label{lem:swmrsa:completeness}
The set of pairs $\mathcal{P}$ returned by an audit in $\pi$ 
satisfies the Completeness property. 
\end{lemma}

\begin{proof}
Consider an audit $\op{a}$ that returns a set $\mathcal{P}$,
and let $\op{r}$ be a read operation by process $p_i$ that returns a value $v$ and that precedes $\op{a}$ in $\pi$. In the following we prove that $(p_i,v)\in \mathcal{P}$.
By Lemma~\ref{lem:swmrsa2}, every read returns the value of the most recent preceding write in $\pi$, denoted $op_w$. 
Let $\op{r}'$ be the first read that returns $v$ to $p_i$. $\op{r}'$ is either $\op{r}$ or it is in between $\op{w}$ and $\op{r}$ in $\pi$. According to the way we linearize the operation, the swap primitive applied during $\op{w}$ precedes the read primitive applied by $p_i$ in the execution of $\op{r}'$. In particular, $\op{w}$ sets $\bit_i$ to $0$ (Line~\ref{line:swmrsa:writeSwap}), and $\op{r}'$ read this bit equal to $0$ and sets its value to $1$ (Line~\ref{line:swmrsa:readFetch}). And this value does not change until the next write operation (if any) applies the swap to write a new value. Moreover, the value in $\currVal$ is $v$ (Line~\ref{line:swmrsa:writeCurVal} ).

If there is no write operation between $\op{r}$ and $\op{a}$, 
then the value read from $R$ when executing the audit operation has $\bit_i$ equals to $1$ (Line~\ref{line:swmrsa:auditCondition}). Also
$\currVal$ is still $v$ when the writer executes Line~\ref{line:swmrsa:auditAddAudit}. 
Thus, $(p_i,v)$ is added to $\mathcal{P}$. 
Otherwise, there is a write between $\op{r}$ and $\op{a}$ in $\pi$.
Let $\op{w}'$ be the first such write, and notice that $\op{w}'$ completes,
since there is a following audit (by the same process). 
Moreover, since it is the first write after $\op{r}$, 
$\bit_i$ is equal to $1$ when the writer executes Line~\ref{line:swmrsa:writeSwap} 
and $\currVal$ is equal to $v$ immediately before it executes Line~\ref{line:swmrsa:writePrevVal}. 
Thus, the pair $(p_i,v)$ is added to $\auditResult$ at Line~\ref{line:swmrsa:auditAddAudit}.
\end{proof}

\begin{lemma}\label{lem:swmrsa:strongAccuracy}
The set of pairs $\mathcal{P}$ returned by an audit in $\pi$ 
satisfies the Strong Accuracy property. 
\end{lemma}

\begin{proof}
Consider an audit operation $\op{a}$ that returns a set $\mathcal{P}$,
and let $(p_i,v)$ be some pair in $\mathcal{P}$. 
The first operation $\op{}$ that adds $(p_i,v)$ to $\mathcal{P}$ is either 
$\op{a}$ itself or a write/audit that precedes $\op{a}$ in $\pi$. 
This is because the variable $\auditResult$ holding the set $\mathcal{P}$ 
is updated after the swap (or read) primitive applied by the writer
in the execution of a write (or audit) operation. 

If $(p_i,v)$ is added to $\mathcal{P}$ by an audit $\op{a}^\prime$, 
then $\currVal$ is $v$ when this pair is added. 
Since condition in Line~\ref{line:swmrsa:auditCondition} holds, 
$\bit_i=1$ when the writer executes Line~\ref{line:swmrsa:auditAddAudit} (since the writer is also the auditor, no write operation can be concurrent to the audit). By Lemma~\ref{lemma:setRead}, only $p_i$ can set $bit_i$ to $1$, then $p_i$ read $v$ and set its bit to $1$ (at Line~\ref{line:swmrsa:readFetch}) before the step of the audit, which proves the lemma.
     
If $(p_i,v)$ is added to $\mathcal{P}$ in the execution of a write $\op{w}$.
By Line~\ref{line:swmrsa:writePrevVal} and~\ref{line:swmrsa:writeAudit}, 
$v$ is the value written by the write operation that immediately precedes 
$\op{w}$ in $\pi$.
Then $v$ is the value returned by the read that set $\bit_i=1$,
which allows the condition in Line~\ref{line:swmrsa:writeSwap} to hold. 
This read precedes $\op{w}$ and therefore, it also precedes $\op{a}$.
\end{proof}

\begin{theorem} \label{theorem:swmrsa:correctness}
Algorithm~\ref{alg:AtomicAuditableRegister:swmrsa} implements
a single-writer multi-reader atomic register 
with single-auditor atomic audit.
\end{theorem}


\subsection{Implementing single-reader atomic register with multi-auditor atomic audit using swap and fetch\&add}
\label{sec:Implementationswsrma}

The algorithm for a single-reader atomic register 
with \emph{multi-auditor} atomic audit follows a similar idea as the 
algorithm for a \emph{multi} reader in the previous section,
using a shared register accessed with the $read$, $swap$ and $fetch\&add$ 
primitives to support the detection of read operations by the writer and the auditors.

Since an audit operation can overlap read, 
write and other audit operations, we need an additional mechanism 
to ensure that the return value of the audit is linearizable. 
The reader and the auditors share information in an unbounded array of read/write 
registers called $pairs$, where $pairs[k]$ indicates whether the reader 
read the $k$-th value written by the writer (if there was such write). 
If $pairs[k]$ contains the initial value $\bot$ then the reader has not read 
the $k$-th value written, otherwise $pairs[k]$ contains that value. 
Each value written has a unique sequence number that is incremented when 
the writer performs a new write.
When performing a write of a value $v$, 
the writer applies a swap to $R$ to atomically write $v$ together 
with its sequence number and set the lowest order bit of the register to $0$ 
to indicate a new write (not yet read). 

Algorithm~\ref{alg:AtomicAuditableRegister:swsrma} presents the pseudo-code.
As in Algorithm~\ref{alg:AtomicAuditableRegister:swmrsa},
in a read operation, the reader reads the value of $R$ 
and sets the low-order bit to $1$ if it was equal to $0$
(indicating that this is the first time $p_r$ read this value).
Additionally, it writes the value read in the corresponding entry of $pairs$. 
When a process performs an audit operation, 
it retrieves from $R$ the sequence number $sn$ of the last write operation,
and also checks whether the reader has read the last value written $v$. 
In the latter case, it writes $v$ into $pairs[sn]$. 
Then, it reads all the entries of $pairs$, from index $sn$ down to the first, 
to obtain its return set. 



Because the value $v$ and the sequence number are unbounded, 
we interleave them bit by bit in $R$, as done in~\cite{NahumABH2022}.
Three functions are used to extract information from $R$.
If $R$ holds $temp=(v,sn,bit)$, then $\mathsf{GetBit}(temp)$ retrieves its lowest-order bit, 
$\mathsf{GetValue}(temp)$ retrieves the value $v$, and $\mathsf{GetSn}(temp)$ retrieves $sn$.

\begin{algorithm}

{ \footnotesize
\begin{algorithmic}[1]

\Statex \textbf{Shared Variables:}
\Statex  $R$ accessed with $\mathit{read}$, $\mathit{fetch\&add}$ and $\mathit{swap}$. Its initial value is $(v_0,0,0)$. 
\Statex $pairs$ An unbounded array of read/write registers, shared by all processes. Initially all registers contains the special value $\perp$.

\Statex
\Statex \textbf{Local Variables:}
\Comment{{\bf Pseudo code for reader $p_r$}}
\Statex $sn$, initially $0$ \Comment{the sequence number of the value store in the register}
\Statex $\readResult$, initially $\perp$ \Comment{value read from the register}

\START[Read()]
    \State{$temp \gets$ $R.read()$} 
    \label{line:swsrma:read}
    \START[if]{ $(GetBit(temp) = 0)$}
    \label{line:swsrma:conditionRead}
        \State{$temp \gets GetValue(\mathit{R.fetch\&add}(1))$}
        \label{line:swsrma:readFetch}
        \State{$\readResult\gets GetValue(temp)$}
        \State{$pairs[GetSn(temp)].write(\readResult)$}
        \label{line:swsrma:lastReadWrite}
    \EndPhase
    \State{\textbf{return} $\readResult$}
    \label{line:swsrma:readReturn}
\EndPhase

\Statex
\Statex \textbf{Local Variables:}
\Comment{{\bf Pseudo code for writer $p_w$}}
\Statex $\prevVal$, initially $\perp$ \Comment{previous value written} 
\Statex $sn$, initially $0$ 
\Comment{the sequence number of the write} 

\START[Write($v$)]
    \State{$sn \gets sn +1$} 
    \label{line:swsrma:snIncrease}
    \State{$temp \gets R.swap((v,sn,0))$}
    \label{line:swsrma:writeSwap} 
    \START[if] $(GetBit(temp) =1)$
    \label{line:swsrma:conditionWriteAudit} 
        \State{$pairs[GetSn(temp)].write(GetValue(temp))$}
        \label{line:swsrma:WritePairsWriter} 
        \Comment{detect the read of the previous write}
    \label{line:swsrma:auditWriterWrite} 
    \EndPhase
    \State{\textbf{return}}
\EndPhase

\Statex
\Statex \textbf{Local Variables:}
\Comment{{\bf Pseudo code for auditor $p_i$}}
\Statex $\mathit{audit\_result}$ initially $\emptyset$ \Comment{set of couples (process, value)} 
\Statex $\mathit{audit\_index}$ initially $0$ \Comment{index of the last updated value in pairs[]} 
\Statex $val$, initially $0$ \Comment{the value store in the register}


\START[Audit()]   
    \State{$temp \gets R.read()$} 
    \label{line:swsrma:auditRead}
    \State{$audit\_index\gets GetSn(temp)$}
    \START[if] {$(GetBit(temp)=1)$}
    \label{line:swsrma:auditConditionAudit}
        \State{$pairs[\mathit{audit\_index}].write(GetValue(temp))$}
 \label{line:swsrma:WritePairsAuditor} 
        \label{line:swsrma:auditWriteAudit}
    \EndPhase
    \START[for]{ $j$ from $\mathit{audit\_index}$ to 0} 
    \label{line:swsrma:auditIteration}
        \START[if]{$(pairs[j].read()\neq \perp)$}    
        \State{$\auditResult.add(p_r, pairs[j].read())$}
        \label{line:swsrma:auditAddAudit}
        \EndPhase
    \EndPhase
    \State{\textbf{return} $\auditResult$}
    \label{line:swsrma:auditReturn}
\EndPhase

\end{algorithmic}
}

\caption{Implementation of a single-reader atomic register with multi-auditor atomic audit using $\mathit{swap}$ and $\mathit{fetch\&add}$}
\label{alg:AtomicAuditableRegister:swsrma}
\end{algorithm}





\subparagraph*{Proof of Correctness:}

We assume that the values written to the register are unique.

Note that there are at most $n$ pending operations in $H$: 
one (either an audit or a write) by the writer, one (either an audit or a read) by the reader,
and possibly one (an audit) for all the other processes.
We construct a history $H'$ by completing some operations in $H$.
We never complete a pending audit.
We complete a pending read in $H$ if and only if some audit contains $(p_r,v)$ 
in its response and no preceding read in $H$ (which must be complete) returns $v$.
After completing the reads, we complete a pending write 
if and only if some (completed) read returns the corresponding value.
We remove from $H'$ all other pending operations in $H$. 
Note that if a pending operation is completed, then it applied a primitive to $R$:
a write is completed if some read has read its value, namely, 
the write has executed the swap in Line~\ref{line:swsrma:writeSwap}; 
a read is completed if it is the only read that returns a value detected by an 
audit, thus, the read has executed the $\mathit{fetch\&add}$ in Line~\ref{line:swsrma:readFetch}.
Thus, all operations in $H'$ applied a primitive to $R$, and we can associate a sequence number $sn$ to each operation, which corresponds to the sequence number they read (for a read or audit operation) or write (for a write operation) from the shared register $R$ during this primitive.

We construct a total order $\pi$ of the operations in $H'$. 
First, we put in $\pi$ all the write operations according to the order they occur in $H'$; because write operations
are executed sequentially by the unique writer, this sequence is well-defined and the order is consistent with the sequence number associated with the values written. 

Next, we add the read operations in $\pi$. Since there is a unique reader the read operations are executed sequentially. The sub-sequence of read operations that returns a value with sequence number $sn$ is placed immediately after the write operation that generates the sequence number $sn$, while preserving their order in $H'$.


The construction of $\pi$ immediately imply that a read operation 
returns the value written by the write preceding it in $\pi$.

\begin{lemma}\label{lem:swsrma:readLastWrite}
Every read operation in $\pi$ returns the value of the most recent preceding write in $\pi$, 
if there is one, and the initial value otherwise. 
\end{lemma}

Finally, 
we consider the audit operations one by one, 
in reverse order of their response in $H$. 
Consider an audit operation $\op{a}$ and let $sn$ be the sequence number it read at Line~\ref{line:swsrma:auditRead}. 
There are three cases. 

\begin{itemize}
    \item Case 1: If $\op{a}$ reads a value $v$ in $pairs[sn]$, 
    we place $\op{a}$ in $\pi$ immediately after the last read 
    with sequence number $sn$ that starts before $\op{a}$ terminates. 
    
    \item Case 2: If $\op{a}$ read a value $v$ in $pairs[sn-1]$ and the initial value $\perp$ in $pairs[sn]$, 
    we place $\op{a}$ in $\pi$ immediately after the write operation with sequence number $sn$ (at the start if no such write exists).

    \item Case 3: If $\op{a}$ read the initial value $\perp$ both in $pairs[sn]$ and in $pairs[sn-1]$, 
    then we place $\op{a}$ in $\pi$ immediately after the write operation with sequence number $sn$ if it terminates before $\op{a}$ is invoked in $H'$. 
    Otherwise, $\op{a}$ is placed immediately after the write operation with sequence number $sn-1$ (at the start if there is no such operation).
\end{itemize}

Case 3 handles the situation where an audit operation $\op{a}$ 
reads a sequence number $sn$ but misses a read operation $\op{r}$ 
that returns the value with sequence number $sn-1$. 
This happens only if $\op{a}$ is concurrent with $\op{r}$ 
and with the write $\op{w}$ that generates the sequence number $sn$; 
in particular, $\op{r}$ and $\op{w}$ write into $pairs[sn-1]$ 
after $\op{a}$ read it.

We next prove that $\pi$ preserves the real-time order of $H$ and 
that the results of the audit operations guarantee the completeness 
and strong accuracy properties.

\begin{lemma}\label{swsrma:realTimeOrder}
    Let $op_1$ and $op_2$ two high-level operations (read, write, and audit) in $H$ 
    such that $op_1$ completes before $op_2$ begins. Then, $op_1$ precedes $op_2$ in $\pi$.
\end{lemma}

\begin{proof}
The construction preserves the order of write-write and read-read operations. 
Consider a write operation $op_w$ and a read operation $op_r$;
let $sn$ be the sequence number generated by $op_w$
and $sn'$ be the sequence number returned by $op_r$.
If $op_r$ precedes $op_w$ in $H'$ then $sn' < sn$,
and by construction $op_r$ is placed after the write that generates $sn'$ 
and before the next write (which is possibly $op_w$). 
If $op_r$ follows $op_w$ in $H$, then $sn' \geq sn$, 
and by construction, $op_r$ is placed after $op_w$ in $\pi$.


Consider now an audit operation $op_a$, and assume 
the sequence number read by $op_a$ (Line~\ref{line:swsrma:auditRead}) is $sn$.

Consider a write operation $op_w$ with sequence number $sn'$.
\emph{(A)} If $op_w$ follows $op_a$ in $H'$, then $sn'\geq sn+1$.
Since the latest $op_a$ is placed in $\pi$ 
is after the write with sequence number $sn$, 
it follows that $op_a$ is ordered before $op_w$ in $\pi$.
\emph{(B)} If $op_w$ precedes $op_a$ in $H'$, 
then $sn' \leq sn$.
Otherwise, $op_a$ would read a value greater than $sn$,
because the sequence numbers are monotonically increasing.
If $op_a$ is ordered after the write operation with sequence number $sn$,
then it is clearly after $op_w$ in $\pi$.
The only case $op_a$ is ordered before the write operation $op_w'$ that generates $sn$ 
is when $op_w'$ does not precede $op_a$ in $H'$, so it cannot be $op_w$. 
Thus, $op_a$ is placed after $op_w$ in $\pi$.


Consider a read operation $\op{r}$ that returns $v$, and let $sn'$ be its sequence number. 
\emph{(A)} If $\op{r}$ follows $\op{a}$ in $H'$, then $sn' \geq sn$.
If $sn' > sn$, then $\op{a}$ is placed before $\op{r}$ in $pi$,
since $\op{a}$ is placed in $\pi$ before the write with sequence number $sn$,
and $\op{r}$ is placed in $\pi$ after the write with sequence number $sn'$.
Otherwise, $sn'=sn$.
If $\op{a}$ reads $v$ from $pairs[sn]$, 
then Case 1 applies, and $\op{a}$ precedes $\op{r}$ in $\pi$. 
If $\op{a}$ reads $\perp$ from $pairs[sn]$, 
then it is placed immediately after the write that generates $sn$ (at the latest)
and thus it precedes $\op{r}$ in $\pi$. 
\emph{(B)} If $\op{r}$ precedes $\op{a}$ in $H$, then $sn' \leq sn$.
If $sn' = sn$, Then $\op{r}$ writes $v$ in $pairs[sn']$ before $op_a$,
and by Case 1, $\op{a}$ is ordered in $\pi$ immediately after the last read 
with sequence number $sn'$ that starts before $\op{a}$ terminates. 
Then $\op{a}$ is ordered after $\op{r}$ in $\pi$.
If $sn' = sn-1$, then since $pairs[sn'] \neq \bot$ when $\op{a}$ reads it,
$\op{a}$ is placed in $\pi$ after the write with sequence number $sn$, 
and hence after $\op{r}$.
Finally, if $sn' < sn-1$ then $\op{r}$ is placed before the write with sequence 
number $sn'-1$, while $\op{a}$ is placed after this write. 


    


Finally, consider another audit operation $\op{a'}$, with sequence number $sn'$.
Without loss of generality, assume $\op{a}$ precedes $\op{a'}$ in $H$;
therefore, $sn \leq sn'$. 
Furthermore, if $\op{a'}$ read $\perp$ from an entry of $pairs$, 
then $\op{a}$ also read $\perp$ in the same entry. 
Thus, $\op{a'}$ is placed after a write with sequence number greater or equal to the last preceding write of $\op{a}$ in $\pi$. Also, if a read operation starts before $\op{a}$ terminates, then it starts before $\op{a'}$ terminates, implying that $\op{a'}$ cannot be linearized before any read operation that precedes $\op{a}$ in $\pi$.
The claim follows because we add audit operations in $\pi$ in reverse order of their responses in $H'$. 
Thus, if they are placed immediately after the same operation (either a read or a write) then $\op{a}$ is placed before $\op{a'}$ in $\pi$.
\end{proof}

\begin{lemma}\label{lem:swsrma:completeness}
The set of pairs $\mathcal{P}$ returned by an audit in $\pi$ 
satisfies the Completeness property. 
\end{lemma}

\begin{proof}
%
%
Consider an audit operation $\op{a}$ that returns a set $\mathcal{P}$, 
and let $\op{r}$ be a read operation by process $p_r$ that 
returns a value $v$ and precedes $\op{a}$ in $\pi$. 
We prove that $(p_r,v) \in \mathcal{P}$.

Let $sn$ be the sequence number read by $\op{a}$, and
let $sn'$ the sequence number of the value read by $\op{r}$.
Since $\op{a}$ follows $\op{r}$ in $\pi$, 
according to our linearization rules $sn\geq sn'$. 
Thus, $audit\_index\geq sn'$ and $\op{a}$ reads $pairs[sn']$ 
(Lines~\ref{line:swsrma:auditIteration}). 
If the read returns $v$, then $(p_r,v)$ is added to $\mathcal{P}$ 
(Line~\ref{line:swsrma:auditAddAudit}), and the lemma follows.
It remains to prove, by way of contradiction, that $\op{a}$
does not read $\perp$ from $pairs[sn']$.
We consider the possible cases:


 \begin{enumerate}
    \item $sn=sn'$, since $\op{a}$ read $\perp$ from $pairs[sn']$, Case 2 or Case 3 apply.
    Thus, $\op{a}$ is placed in $\pi$ before the write that generates 
    sequence number $sn'$ (at the latest). 
    Since $\op{r}$ follows this write, $\op{a}$ is placed before $\op{r}$ in $\pi$.
    

    \item $sn=sn'+1$: 
    Case 2 does not hold because $\op{a}$ read $\perp$ from $pairs[sn']$ with $sn'=sn-1$. 
    Since $\op{r}$ precedes $\op{a}$ in $\pi$, by Case 3, 
    $op_a$ follows the write with sequence number $sn$ in $\pi$.
    There are two cases: 
   
    \begin{enumerate}
    
        \item Case 1: $\op{a}$ read a value $\neq \perp$ from $pairs[sn]$. 
        We reach a contradiction by showing $\op{a}$ cannot read $\perp$ from $pairs[sn-1]$. 
        Since $pairs[sn]=v'$, there is a read operation $\op{r'}$ that reads $v'$, 
        and since there is a single reader, $\op{r'}$ follows $\op{r}$ in $H$. 
        Thus, the value of $pairs[sn-1]$ is set to $v$ before $pairs[sn]$ is set to $v'$. 
        By Line~\ref{line:swsrma:auditIteration}, 
        $\op{a}$ first reads $pairs[sn]$ and then reads $pairs[sn-1]$,
        and therefore, it does not read $\perp$ from $pairs[sn-1]$. 
        

        \item Case 3: $\op{a}$ read $\perp$ from $pairs[sn]$, 
        but the write $\op{w}$ that generates $sn$ precedes $\op{a}$. 
        Since $\op{r}$ returns $v$, there is a read operation (possibly $\op{r}$) 
        that set the corresponding low-order bit to $1$ (Line~\ref{line:swsrma:readFetch}). 
        Then, when $\op{w}$ writes the new value with sequence number $sn$, 
        it reads $1$ from this bit and sets $pairs[sn-1]=v$,
        so $\op{a}$ does not read $\perp$ from $pairs[sn-1]$.
        
 \end{enumerate}
    \item $sn \geq sn'+2$. 
    Then the write operation with sequence number $sn'+1$ completes before $\op{a}$ reads $R$,
    and we can apply the same reasoning as in case 2.b. 

 \end{enumerate}
\end{proof}

\begin{lemma}\label{lem:swsrma:strongAccuracy}
The set of pairs $\mathcal{P}$ returned by an audit in $\pi$ 
satisfies the Strong Accuracy property. 
\end{lemma}

\begin{proof}
Consider an audit operation $\op{a}$ that returns a set $\mathcal{P}$,
and let $(p_r,v)$ be a pair in $\mathcal{P}$. 
We prove that a read operation $\op{r}$ by process $p_r$ that returns $v$
is placed before $\op{a}$ in $\pi$.

Since $(p_r,v)$ is in $\mathcal{P}$, 
$\op{a}$ adds $(p_r,v)$ to $audit\_result$ 
because it read $pairs[sn]=v$ for some $sn$. 
If $v$ was written by the reader (Line~\ref{line:swsrma:lastReadWrite}), 
then the reader returned the value $v$ associated with $sn$, 
in Line~\ref{line:swsrma:readFetch}. 
Otherwise, $v$ was written to $pairs[sn]$ either 
by the writer (Line~\ref{line:swsrma:WritePairsWriter})
or by the auditor (Line~\ref{line:swsrma:WritePairsAuditor}). 
This means that the writer or the auditor read the bit set to $1$ 
when, respectively, checking the condition in Line~\ref{line:swsrma:conditionWriteAudit} 
or in Line~\ref{line:swsrma:auditConditionAudit}. 
This bit is set to $1$ only by the reader when reading the 
corresponding value in Line~\ref{line:swsrma:readFetch}. 

Thus, there is a read operation $\op{r}$ that read the value $v$ with sequence number 
$sn$ that does not follow $\op{a}$ in $H$. 
We now show that it precedes $\op{a}$ also in $\pi$.

Let $sn'$ be the sequence number of of $\op{a}$.
Since $\op{a}$ reads $pairs[sn]$, $sn' \geq sn$.
If $sn'=sn$, then since $\op{a}$ read $v$ in $pairs[sn]$, 
$\op{a}$ is placed after $\op{r}$ in $\pi$ and the claim holds.
If $sn' = sn+1$, then since $\op{a}$ reads $v$ from $pairs[sn]$, 
by Cases 1 and 2, it is placed after the write that generates 
sequence number $sn+1$, and therefore, after $\op{r}$. 
Finally, if $sn'> sn+1$, then $\op{a}$ is placed in $\pi$ after a write with
a sequence number greater than $sn$, and hence, after $\op{r}$. 
\end{proof}

\begin{theorem} \label{theorem:swsrma:correctness}
Algorithm~\ref{alg:AtomicAuditableRegister:swsrma} implements
a single-writer single-reader atomic register 
with multi-auditor atomic audit.
\end{theorem}


\subsection{Implementing multi-reader atomic register with multi-auditor atomic audit using compare\&swap}
\label{sec:casAlgorithm}


To deal with multiple readers, as in Algorithm~\ref{alg:AtomicAuditableRegister:swmrsa},
each reader sets a dedicated bit in the $n$ lower-order bits of a shared register $R$ 
and the writer writes the value together with a sequence number in the higher-order bits of $R$. 
To deal with multiple auditors, we use an array $\mathit{pairs}$, 
as in Algorithm~\ref{alg:AtomicAuditableRegister:swsrma}. 
To accommodate multiple readers, the array is bi-dimensional, 
with an unbounded number of columns (corresponding to each written value) 
and $n$ rows, one for each reader. 
Specifically, $\mathit{pairs}[i][sn]=\perp$ indicates that process $p_i$ 
has not read the value $v$ written by the $sn$-th write operation (if any);
otherwise, $\mathit{pairs}[i][sn]=v$.


We do not need readers to write into pairs because the writer applies a $\mathit{compare\&swap}$ 
to write the new value in $R$, using the previously-read state of $R$. 
So, if in the meanwhile, some reader read the current value stored in $R$ 
and set its bit to $1$, the $\mathit{compare\&swap}$ fails. 
Thus, the writer detects and write into $pairs$ all the read operations 
of the last value written before succeeding the next write. 
Thus, either the auditor can detect the read operations by reading $R$ 
because the bits were not reset by the new write, or this information is in $pairs$.



Because the sequence numbers and the corresponding values are unbounded, 
we cannot a priori divide the high-order bits between them. 
Instead, we interleave them bit-by-bit, 
as done in Algorithm~\ref{alg:AtomicAuditableRegister:swsrma}
(following~\cite{NahumABH2022}).

Each process has a local variable $\temp$ to store the value read from $R$,
in order to select the information it needs.
For example, a read operation by a process $p_i$ has to retrieve the value 
of the low order bit associated with $p_i$ to check if it is equal to $1$ or $0$, 
meaning it has already read the current value or not, respectively. 
Given a value $\temp$, we use the following functions:
$\mathsf{GetValue}(\mathit{temp})$ retrieves 
the value stored in the high-order bits of $\temp$, 
$\mathsf{GetSn}(\mathit{temp})$ retrieve the 
sequence number stored in the high-order bits of $\temp$,
and $\mathsf{GetBits}(\mathit{temp})$ retrieves the $n$ low-order bits of $\temp$. 

A reader has a local variable called $\mathit{read\_result}$ to store the value 
that has to be returned, initially $\perp$. 
This is used to ensure that if a new value was not written, 
consecutive read operations by the same process can return the correct value 
without setting the corresponding bit to $1$ more than once. 

An auditor has a local variable called $\mathit{audit\_result}$ 
that holds a set of pairs $(process,value)$, 
one for each detected read operation; it is initially $\emptyset$. 
The local variable $\mathit{audit\_index}$ holds the sequence number read from 
$R$, indicating the last column of the matrix $\mathit{pairs}$ 
written by the writer that the auditor has to read; it is initially $0$.

The writer holds the last value written in a local variable $\mathit{val}$.

The pseudocode appears in Algorithm~\ref{alg:AtomicAuditableRegister:swmrma}.
\begin{algorithm}
{ 
\begin{algorithmic}[1]
\Statex \textbf{Shared Variables:}

\Statex $\mathit{R}$: a register shared by all processes, accessed with $read$, $compare\&swap$, and $fetch\&add$. 
\Statex It contains a sequence number, the corresponding value, and $n$ bits. Initially ($0,v_0,0^n$).

\Statex $\mathit{pairs[n,\ldots]}$: 
a matrix of read/write registers, where $\mathit{pairs[j][k]}$ indicates if process $p_j$ has read the k-th written value. Initially, $\perp$. 
\Statex 

\Statex \textbf{Local Variables:} \Comment{{\bf Pseudo code for reader and auditors $p_i$ ($i\in[0,n-1]$)}}
\Statex $\mathit{temp}$ initially $\perp$ \Comment{the content of the register $R$} 
\Statex $\mathit{read\_result}$ initially $\perp$ \Comment{the last value read}
\Statex $\mathit{audit\_result}$ initially $\emptyset$ \Comment{set of (process, value) pairs} 
\Statex $\mathit{audit\_index}$ initially $0$ \Comment{index of the last updated value in pairs[]} 

\Statex

\START[Read()] 
    \State{$temp \gets \mathit{R.read()}$}
    \label{line:swmrma:read}
    \START[if]{ $(\mathsf{GetBits}(temp)[i] = 0)$}
    \label{line:swmrma:readCondition}
    \State{$\mathit{read\_result}  \gets \mathsf{GetValue}(\mathit{R.fetch\&add}(2^i))$}
    \label{line:swmrma:readFetch}
    \EndPhase
    \State{\textbf{return} $\mathit{read\_result}$}
    \label{line:swmrma:readReturn}
\EndPhase

\Statex

\START[Audit()]
    \State{$\mathit{temp} \gets \mathit{R.read()}$}
    \label{line:swmrma:auditRead}
    \State{$\mathit{audit\_index} \gets \mathsf{GetSn}(temp)$}
     \label{line:swmrma:auditIndex}
    \START[for]{ $0\leq j < n$}
        \START[for]{ $0\leq k < \mathit{audit\_index}$}
            \START[if]{$(pairs[j][k].read()\neq \perp)$}    
                \State{$\auditResult.add(p_j, pairs[j][k].read())$}
                \label{line:swmrma:auditAddAuditPairs}
                \EndPhase
            \EndPhase
        \EndPhase
        
    \START[for] $0 \leq j < n$
            \START[if] $(\mathsf{GetBits}(temp)[j] = 1)$
                \Comment{checks if $p_j$ read the last value written in R}
                \label{line:swmrma:auditCondition}
                \State{$\mathit{audit\_result}.add(p_j,\mathsf{GetValue}(temp))$}
                \label{line:swmrma:auditAddAudit}
            \EndPhase
    \EndPhase

    \State{\textbf{return} $\mathit{audit\_result}$}      
\EndPhase

\Statex

\Statex \textbf{Local Variables:} \Comment{{\bf  Pseudo code for writer $p_0$}}
\Statex $\mathit{temp}$ initially $\perp$ \Comment{the value read from $R$}
\Statex $\mathit{sn}$ initially $0$ \Comment{sequence number of the high-level writes} 
\Statex $\mathit{val}$ initially $v_0$ \Comment{input value of the last high-level write}
\Statex $\mathit{bits[]}$ initially $0^n$ \Comment{$n$ lowest-order bits of $R$ to detect high-level reads}

\Statex

\START[Write($v$)]
    \State{$sn \gets sn+1$}
     \label{line:swmrma:seqNum}
    \START[while]{$(\mathit{R.compare\&swap}((\mathit{sn-1},\mathit{val},bits),(\mathit{sn},v,0^n))\neq \mathit{True})$}
     \label{line:swmrma:writeSwap}
        \State{$\mathit{temp} \gets \mathit{R.read()}$}
        \label{line:swmrma:writeRead}
        \State{$\mathit bits \gets \mathsf{GetBits}(temp)$}
        \START[for] $0 \leq j < n$
            \label{line:swmrma:loopAudit}
            \START[if] $(bits[j] = 1)$
             \Comment{check if $p_j$ read the last value}
                \label{line:swmrma:writecheckone}
                \State{$\mathit{pairs[sn-1][j].write(val)}$} 
                \label{line:swmrma:writeAudit}

            \EndPhase
        \EndPhase  
    \EndPhase
    \State{$bits \gets 0^n$}
    \State{$val\gets v$}
    \State{\textbf{return}}
\EndPhase

\end{algorithmic}
}

\caption{Implementation of a multi-reader atomic register with multi-auditor atomic audit using $\mathit{compare\&swap}$ and $\mathit{fetch\&add}$ for $n$ processes.}
\label{alg:AtomicAuditableRegister:swmrma}
\end{algorithm}

In a read, the reader first reads $R$ to check whether 
it has already read its last value.  
If this is the case, it simply returns the value.
Otherwise, it applies a $\mathit{fetch\&add}$ to set its bit to $1$
(indicating that it read the value) and returns the value represented 
by the high-order bits of the value returned by the $\mathit{fetch\&add}$. 

In an audit, the auditor first reads $R$ to get the sequence of bits 
indicating the read operations performed since the last write operation
and to atomically get the sequence number of the last write operation. 
It then reads all the pairs stored in entries of $\mathit{pairs}$ until this index,
and adds them to the set that the audit will return; this set is persistent. 
Finally, it adds to the set additional pairs corresponding 
to the low-order bits of $R$ that are set.
For simplicity, all the audit reads all entries of $pairs$ starting from $0$ 
up to the read sequence number. 
It is simple to ensure that the same auditor reads each column of $pairs$ at 
most once, by using a persistent local variable to store the last column read.

To write a new value $v$, the writer increments the sequence number $sn$, 
and applies $\mathit{compare\&swap}$ to $R$ to store $sn$ together with $v$ 
and reset the $n$ low-order bits to 0.
If this is successful, the operation completes; 
otherwise, the writer reads $R$, 
and for each of the $n$ low-order bits that is set to $1$, 
it writes $v$ into the corresponding entry of $\mathit{pairs}[\,][sn-1]$ 
to announce the read operation that set the bit.
The writer then retries the  $\mathit{compare\&swap}$.



\subparagraph*{Proof of Correctness:}
We assume that the values written to the register are unique.
We first show that all operations (by a correct process) complete within a finite number of steps. 
This is immediate for read operations. 
For an audit operation, the number of the iterations of the first $\mathit{for}$ loop 
is bounded by the value $index$ read from $\mathit{R}$,
while the second $\mathit{for}$ loop has $n$ iterations.
We next bound the number of iterations in a write operation.

\begin{lemma}\label{lemma:swmrma:setRead}
For every $i \in\{0,\ldots, n-1\}$, 
$\bit_i$ is set to $0$ each time a write operation successfully executes 
the $\mathit{compare\&swap}$ at Line~\ref{line:swmrma:writeSwap}, 
and it is set to $1$ if and only if a read operation by $p_i$ executes 
Line~\ref{line:swmrma:readFetch}.
\end{lemma}

\begin{proof}
When a write operation executes Line~\ref{line:swmrma:writeSwap}, 
it resets the $n$ least significant bits of $R$ to $0$,
together with the new value.
For each $i \in\{0,\ldots n-1\}$, 
process $p_i$ adds $2^{i}$ to the value stored in $R$ if and only if 
the $i$-th bit of its binary representation is $0$ 
(Line~\ref{line:swmrma:readCondition}). 
Thus, $p_i$ sets the $i$-th bit to $1$ and does not change any other bit of $R$.
\end{proof}

\begin{lemma}\label{lemma:swmrma:WaitFree}
The $\mathit{compare\&swap}$ (Line~\ref{line:swmrma:writeSwap}) of a write operation 
fails at most $n$ times.
\end{lemma}

\begin{proof}
The $\mathit{compare\&swap}$ fails if the state of $R$ changes in between 
the $read$ (Line~\ref{line:swmrma:writeRead}) and 
the following $\mathit{compare\&swap}$ (Line~\ref{line:swmrma:writeSwap}). 
Only a reader can change the state of $R$ between the application of these 
two primitives by the writer.
By Lemma~\ref{lemma:swmrma:setRead}, $\bit_i$ can be set to $1$ in Line~\ref{line:swmrma:readFetch} by process $p_i$ and it is not reset to $0$ unless a new high-level write is performed.
By Line~\ref{line:swmrma:readCondition}, each reader sets its bit at most once 
for each given value, which completes the proof.
\end{proof}

To prove the linearizability of the algorithm, fix a history $H$.
Note that there are at most $n$ pending operations in $H$,
one for each process. 
We construct a history $H'$ by completing some pending 
read and write operations in $H$;
we never complete a pending audit. 
We first complete a pending read invoked by process $p_i$ in $H$ 
if and only if some audit contains $(p_i,v)$ 
in its response and no read in $H$ (which must be complete) returns $v$ to $p_i$.
After completing the reads, we complete a pending write 
if and only if some (completed) read returns the corresponding value.
We remove from $H'$ all other pending operations in $H$. 

Note that if a pending operation is completed, then it applied a primitive to $R$:
A read is completed if it is the only read that returns a value detected by an 
audit, thus, the read has executed $\mathit{fetch\&add}$ in Line~\ref{line:swmrma:readFetch}.
A write is completed if some read has read its value, namely, 
the write has executed the $\mathit{compare\&swap}$ in Line~\ref{line:swmrma:writeSwap}.

We construct a sequential history $\pi$ that contains all the operations in $H'$,
while preserving their real-time order.
We (totally) order of all the read, audit and write operations in $H'$ 
according to the order they apply their last primitive on $R$: 
this is either a $\mathit{read}$ or a $\mathit{fetch\&add}$ for read operations, 
it is a $\mathit{read}$ for an audit, 
and the $\mathit{compare\&swap}$ for write operations.
Note that these are atomic primitives and their order is well-defined. 
Clearly, 
operations are linearized inside their execution intervals,
implying that $\pi$ preserves the real-time order of all the operations in $H$.
Furthermore, we have:

\begin{lemma}\label{lem:swmrma21}
A read in $\pi$ returns the value of the most recent preceding write in $\pi$, 
if there is one, and the initial value, otherwise. 
\end{lemma}
    
\begin{proof}
Consider a read $\op{r}$ by a process $p$ that returns a value $v$. 

If the condition in Line~\ref{line:swmrma:readCondition} holds, 
then the result of $\op{r}$ is the value extracted from the value 
read by applying $\mathit{fetch\&add}$ on $R$ 
in Line~\ref{line:swmrma:readFetch}. 
Thus, there is a preceding successful $\mathit{compare\&swap}$ primitive
that 
writes $v$ into $R$, in the execution of a write operation $\op{w}$.
According to our linearization rules, $\op{w}$ precedes $\op{r}$ in $\pi$. 
This also shows that $\op{w}$ is the most recent preceding write in $\pi$,
since otherwise, $\op{r}$ would have returned a different value.

If the condition in Line~\ref{line:swmrma:readCondition} does not hold, 
then the value returned by $\op{r}$ has been obtained by a previous 
read operation by $p$, for which 
the condition of Line~\ref{line:swmrma:readCondition} held, 
denoted $\op{r'}$. 
By Lemma~\ref{lemma:swmrma:setRead} and since $\bit_i\neq 0$, 
no write changed the values of $R$ after $\op{r'}$ and before $\op{r}$. 
Thus, the claim follows from the previous case.
\end{proof}

\begin{lemma}\label{lem:swmrma:strongAccuracy}
The set of pairs $\mathcal{P}$ returned by an audit in $\pi$ 
satisfies the Strong Accuracy property. 
\end{lemma}

\begin{proof}
Consider an audit operation $\op{a}$ by a process $p$ that returns a set $\mathcal{P}$,
and let $(p_i,v)$ be a pair in $\mathcal{P}$. We prove that there is a read operation by process $p_i$ that returns the value $v$ and it is linearized before $\op{a}$. 
Assume that $\op{a}$ is the first audit operation by $p$ that adds $(p_i,v)$ to $\auditResult$. If not, we can consider the first such audit operation by $p$ and since it has to be linearized before $\op{a}$, the claim follows.

First, consider that $\op{a}$ adds $(p_i,v)$ to $\auditResult$ at Line~\ref{line:swmrma:auditAddAudit}. Then, the condition at Line~\ref{line:swmrma:auditCondition} is verified and the value read from $R$ (Line~  \ref{line:swmrma:auditRead}) has $val=v$ and $\bit_i=1$.
Then, by Lemma~\ref{lemma:swmrma:setRead}, there is a read operation $\op{r}$ that set $\bit_i$ to 1, and that read the value $v$ by applying the $\mathit{fetch\&add}$ to $R$ in Line~\ref{line:swmrma:readFetch}.
Furthermore, the $\mathit{fetch\&add}$ is applied before $\op{a}$ read $R$ in Line~\ref{line:swmrma:auditRead}. 
Thus, $\op{r}$ is linearized before $\op{a}$.

Otherwise, $(p_i,v)$ is added to $\auditResult$ because $p$ 
read $v$ from $\mathit{pairs}[i][k]$, $k< audit\_index$ 
(Line~\ref{line:swmrma:auditAddAuditPairs}). 
Then, the writer has written $v$ into $\mathit{pairs[i][k]}$ 
(Line~\ref{line:swmrma:writeAudit}) in the execution of the write operation 
$\op{w}$ with sequence number $k+1$ (Line~\ref{line:swmrma:writeAudit}).
This means that the condition of Line~\ref{line:swmrma:writecheckone} holds, 
and process $p_i$ previously applied Line~\ref{line:swmrma:readFetch} 
in a read operation that returns $v$;
furthermore, this was before $\op{w}$ applied its successful $\mathit{compare\&swap}$.
Then, $\op{r}$ precedes $\op{w}$ in $\pi$.
Since there is a single writer, $\op{w}$ is either the write that generates 
the sequence number in $audit\_index$, or it is linearized before it. 
This completes the proof since $\op{a}$ is linearized after $\op{w}$. 
%
%
%
\end{proof}

\begin{lemma}\label{lem:swmrma:completeness}
The set of pairs $\mathcal{P}$ returned by an audit in $\pi$ 
satisfies the Completeness property. 
\end{lemma}

\begin{proof}
Consider an audit operation $\op{a}$ that returns a set $\mathcal{P}$,
and let $\op{r}$ be a read operation by process $p_i$ that returns a value $v$.
We prove that if $\op{r}$ is linearized before $\op{a}$ in $\pi$,
then $(p_i,v)\in \mathcal{P}$.

By Lemma~\ref{lem:swmrma21}, 
a read returns the value of the last write that precedes it in $\pi$, 
denoted $\op{w}$. 
Let $\op{r}^{\prime}$ be the first read that returns $v$ to $p_i$;
it is either $\op{r}$ itself or it is linearized between $\op{w}$ and $\op{r}$ in $\pi$.
By Lemma~\ref{lemma:swmrma:setRead},
$\op{w}$ sets $\bit_i$ to $0$ (Line~\ref{line:swmrma:writeSwap}), 
and $\op{r}^{\prime}$ read this bit equal to $0$ 
and sets it to $1$ (Line~\ref{line:swmrma:readFetch}).
This bit and the value $v$ does not change until the next $\mathit{write}$ (if any) applies 
$\mathit{compare\&swap}$ to $R$. 

First, assume there is no write between $\op{r}$ and $\op{a}$. 
Then, by the linearization of read and audit operations, 
the value read from $R$ (in Line~\ref{line:swmrma:auditRead})
contains $\bit_i=1$ and the value $v$. 
Thus, the condition in Line~\ref{line:swmrma:auditCondition} 
is satisfied and $(p_i,v)$ is added to $\mathcal{P}$ 
(in Line~\ref{line:swmrma:auditAddAudit}). 

Otherwise, there is a write between $\op{r}$ and $\op{a}$ in $\pi$.
Let $\op{w}^{\prime}$ be the first such write. 
Since $\op{w}^{\prime}$ is linearized between 
$\op{r}$ and $\op{a}$ in $\pi$,
$\op{w}^{\prime}$ successfully executes the $\mathit{compare\&swap}$ 
before $\op{a}$ reads $R$ (in Line~\ref{line:swmrma:auditRead}).
%
The sequence number of $\op{w}^{\prime}$ is $sn+1$ and 
$\op{a}$ read a sequence number greater than or equal to $sn+1$. 
Thus, $\op{a}$ reads $\mathit{pairs}[i][sn]$ (Line~\ref{line:swmrma:loopAudit}),
and we argue that the read returns $v$, 
implying it adds $(p_i,v)$ to $\mathcal{P}$ (Line~\ref{line:swmrma:auditAddAuditPairs}). 
By Lemma ~\ref{lemma:swmrma:setRead}, the condition of 
Line~\ref{line:swmrma:auditCondition} holds for $bits[i]$
and the writer writes the value written by the last completed write, $v$,
into $\mathit{pairs}[i][sn]$. 
\end{proof}


\begin{theorem} \label{theorem:swmrma:correctness}
Algorithm~\ref{alg:AtomicAuditableRegister:swmrma} implements
a single-writer multi-reader atomic register 
with multi-auditor atomic audit.
\end{theorem}

Note that all (process,value) pairs must be stored somewhere in order to allow 
the audit to return all the read values.
When there is a single auditor, 
as in Section~\ref{sec:Implementationswsrsa} 
and Section~\ref{sec:Implementationswmrsa},
the pairs are stored locally at the auditor. 
This space can be reduced if the completeness property is weakened 
to require that the audit operation returns 
only the last $k \geq 1$ values read by each reader. 
This weaker form of completeness does not affect 
the consensus number.

\section{The consensus number of atomic audit}
\label{sec:swsrsaConsensus2}

The \emph{consensus number}~\cite{H91} of 
a concurrent object type $X$
is the largest positive integer $m$ such that 
consensus can be wait-free implemented from any number of read/write registers, 
and any number of objects of type $X$, 
in an asynchronous system with $m$ processes.
If there is no largest $m$, 
the consensus number is infinite. 

By Theorem~\ref{theorem:async:consensusTwoProcesses},
it is possible to solve consensus among two processes, 
using only swsr atomic registers with single-auditor atomic audit. 
This implies that the consensus number of a swsr atomic register 
with single-auditor atomic audit is at least 2. 
Clearly, the same holds if the register is multi-reader
or multi-auditor. 

To prove that the consensus number of this object type is 2, 
we provide algorithms that implement it with a single auditor 
(with single or multiple readers),
using a single register that supports a combination of read, swap and $fetch\&add$. 
In particular, Algorithm~\ref{alg:AtomicAuditableRegister:swsrsa} implements 
a swsr atomic register with a single-auditor atomic audit  by applying swap and read primitive operations on a single register. Algorithm~\ref{alg:AtomicAuditableRegister:swmrsa} implements a 
multi-reader atomic register with a single-auditor atomic audit 
by applying swap, fetch\&add, and read primitives on a single register.

Our result holds since Herlihy~\cite{H91} proves that one cannot construct a 
wait-free solution of three process consensus using registers that 
support any combination of read, write, swap and $fetch\&add$. 
It follows that a swsr atomic register with a single-auditor atomic audit cannot 
be used to solve consensus among three or more processes, which implies:




\begin{proposition}
\label{theorem:async:swsrsa:ConsensusNumberTwo}
A single-writer 
multi-reader atomic register 
with a single-auditor atomic audit has consensus number two.
\end{proposition}

Similarly, Algorithm~\ref{alg:AtomicAuditableRegister:swsrma} implements 
a single-reader atomic register with multi-auditor atomic audit. 
It only uses a register accessed by read, swap and fetch\& add primitives and 
an unbounded number of read/write registers. 
Thus, the above Herlihy's impossibility together 
with our Theorem~\ref{theorem:async:consensusTwoProcesses}, imply: 

\begin{proposition}
\label{theorem:async:swsrma:ConsensusNumberTwo}
A single-writer single-reader atomic register 
with multi-auditor atomic audit has consensus number two.
\end{proposition}

We also show that a multi-reader atomic register with multi-auditor atomic audit 
has consensus number larger than $2$ if each reader is also an auditor of the register. 
In particular, according to Algorithm~\ref{alg:async:consensusInfinite}, 
we can solve consensus among $n$ processes using $n$-reader 
atomic registers with $n$-auditors atomic audit. 
Thus, the consensus number of this object type is at least $n$.

We finally provide an implementation of the swmr atomic register with multi-auditor 
atomic audit using read/write registers and a register accessed via read, 
fetch\&add and compare\&swap primitives. 
Even though, an object type that supports these three primitives is not traditionally 
used, current architectures support this combination of primitives. 
Since compare\&swap has consensus number infinite, 
it is an open question whether the consensus number of an $n$-reader 
$n$-auditor atomic register with atomic audit is $n$ or more.
Specifically, the question is whether it can be implemented from 
objects with consensus number $n$, e.g., 
similarly to the {\sf Proof} object of~\cite{FreyGR2023}.



Table~\ref{table:atomic:consensusNumber} summarize the results.
\begin{table}[htbp]
\centering
\caption{Consensus number of atomic register with atomic audit}

\begin{tabularx}{\textwidth}{ 
  | >{\centering\arraybackslash}X 
  | >{\centering\arraybackslash}X 
  | >{\centering\arraybackslash}X 
  | >{\centering\arraybackslash}X 
  | >{\centering\arraybackslash}X | }
\hline
Audit consistency & Number of writers & Number of readers & Number of auditors & Consensus number  \\
\hline
Atomic & 1 & 1 & 1 & 2  \\
\hline
Atomic & 1 & $n$ & 1 & 2  \\
\hline
Atomic & 1 & 1 & $n$ & 2 \\
\hline
Atomic & 1 & $n$ & $n$ & $\geq n$ \\
\hline
\end{tabularx}

\label{table:atomic:consensusNumber}
\end{table}


\section{Regular Audit}\label{sec:reg audit imp}
In this section, we show how to implement a multi-reader atomic register with 
multi-auditor regular audit using only single-writer multi-reader atomic registers.
The implementation follows a straightforward approach:
during a read operation each reader leaves a trace in a register 
of all the values they read. 

The pseudo-code appears in Algorithm~\ref{alg:AuditableAtomicRegister:swmrma},
and it uses several atomic registers.
A swmr atomic register $R_v$ is shared between the writer and the readers. 
This register is used by the writer to write a new value and by the readers to access it. 
In addition to $R_{v}$, each reader $p_i$ shares a swmr atomic register $R_{a}[i]$ with the auditors. 
This register is used by $p_i$ to communicate to the auditor 
all the values it read from the register.

In a read, a reader $p_i$ reads a value from $R_{v}$ and stores 
it in $\readValueRegMA$. 
Then, it adds $v$ together with its identifier in $\readLogRegMA$, 
and writes $\readLogRegMA$ in the register $R_{a}[i]$ it shares with the auditors. 
Finally, it returns $\readValueRegMA$.
In a write, the writer simply writes $v$ in $R_{v}$. 
In an audit, the auditor simply reads from all the swmr register $R_{a}$ of each reader, 
combines it with the result in $\auditResult$, and returns. 

\begin{algorithm}[tb]

{
\begin{algorithmic}[1]
\Statex \textbf{Shared Variables:}
\Statex  $R_{v}$, swmr atomic register, initially $v_0$
\Statex {$\forall i \in [0,\Nreader-1]$} 
    $R_{a}${[i]} swmr atomic register with writer $p_j$, 
    readers the auditors. Initially $\perp$

\Statex

\Statex \textbf{Local Variables:}
\Comment{{\bf Pseudo code for reader $p_i$}}
\Statex $\readLogRegMA$, initially $\emptyset$ \Comment{tuples $(p_j, v)$, for each value $v$ read by $p_j$}
\Statex $\readValueRegMA$, initially $\perp$ \Comment{value the reader read}
\START[Read()]
    \State{$\readValueRegMA \gets R_{v}.read()$}
    \label{line:reg:swmrma:readReg}
    \State{$\readLogRegMA.add(p_i, \readValueRegMA)$}
    \label{line:reg:swmrma:readLog}
    \State {$R_{a}{[i]}.write(\readLogRegMA)$}
    \label{line:reg:swmrma:readAudit}
    \State {\textbf{return} $\readValueRegMA$}
    \label{line:reg:swmrma:readReturn}
\EndPhase

\Statex

\START[Write($v$)]
\Comment{Pseudo code for writer $p_0$}
    \State {$R_{v}.write(v)$} 
    \label{line:reg:swmrma:write}
\EndPhase

\Statex

\Statex{\textbf{Local Variables:}}
\Comment{{\bf Pseudo code for auditor $p_k$}}
\Statex 
$\auditResult$, initially $\emptyset$ \Comment{tuples $(p,v)$, with $p$ the reader and $v$ the value.}
\START[Audit()]
     \START[{for}] {$1\leq j< \Nreader$}
    \State {$\auditResult.add$}$(R_{a}{[j]}.read())$
    \label{line:reg:swmrma:AuditAddRead}
    \EndPhase
    \State{\textbf{return} {$\auditResult$}}
    \label{line:reg:swmrma:AuditReturn}
\EndPhase

\end{algorithmic}
}

\caption{Implementation of a single-writer multi-reader atomic register with
\emph{multi-auditor regular audit} using only read and write}
\label{alg:AuditableAtomicRegister:swmrma}
\end{algorithm}

\subparagraph*{Proof of Correctness:}
We assume that the values written to the register are unique.

A history $H$ has at most $n$ pending operations, one for each process.
We never complete a pending audit.
We complete a pending read invoked by process $p_i$ in $H$ 
if and only if some audit contains $(p_i,v)$ 
in its response and no read in $H$ (which must be complete) 
returns $v$ to $p_i$.
Any pending operation that is completed has accessed $R_{v}$
(with a write or a read): 
A read is completed if it is the only read that returns a value detected by an 
audit, the read has executed the $\mathit{write}$ at Line~\ref{line:reg:swmrma:readReg}.
A write is completed only if some read has read its value, namely, 
the write has executed the $\mathit{write}$ in Line~\ref{line:reg:swmrma:write}. 

We totally order all the completed operations by the 
order they access $R_{v}$ (in a $\mathit{read}$ or a $\mathit{write}$).
Since each operation accesses $R_{v}$ exactly once, and since this is 
a primitive atomic operation, the order is well-defined. 
Call this total order $\pi$ and note that it respects the real-time order 
of the high-level read and write operations on the register, since the steps used to 
create it are in the execution interval of the corresponding operation. 

The algorithm uses a shared variable $R_v$ that holds a value along with an array of $n$ registers of unbounded size.
Instead of using an array of $n$ registers of unbounded size, it is quite simple to adapt the solution and uses, similarly to Algorithm~\ref{alg:AtomicAuditableRegister:swmrma}, an array of $m$ registers, 
where $m$ is the number of write operations;
each register containing a pair (process, value).
Furthermore, as discussed for Algorithm~\ref{alg:AtomicAuditableRegister:swmrma}, 
the space complexity can be reduced if the completeness property is weakened
to require that the audit operation returns only the last $k \geq 1$ 
values read by each reader. 
This allows to reduce the space complexity to a single register that holds a value along with an array of $n$ bounded registers.

Note that all the loops in the write or audit operations are iterated at most $n$ times,
and hence Algorithm~\ref{alg:AuditableAtomicRegister:swmrma} is wait-free. 

Since the $\mathit{read}$ primitives 
(Line~\ref{line:reg:swmrma:readReg} and Line~\ref{line:reg:swmrma:readReturn}) 
and the $\mathit{writes}$ primitives (Line~\ref{line:reg:swmrma:write}) 
are applied on an atomic register $R_v$,
we have:
   
\begin{lemma}\label{lem:reg:swmrma2}
    Every read in $\pi$ returns the value of the most recent preceding write, 
    if there is one, and the initial value, otherwise.
\end{lemma}

\begin{lemma}\label{lem:reg:swmrma:completeness}
The set of pairs $\mathcal{P}$ returned by an audit  
satisfies the Completeness Property.
\end{lemma}

\begin{proof}
A pending read $\op{r}$ is completed only if some audit operation 
that detects it, 
and the lemma follows from the construction of $\sf{complete}(H)$.
If $\op{r}$ returns $v$ (Line~\ref{line:reg:swmrma:readReturn}), 
then it adds $(p,v)$ to $\readLogRegMA$ before returning 
(Line~\ref{line:reg:swmrma:readLog}). 
By the code, $(p,v)$ is never removed from $\readLogRegMA$.
Therefore, $R_a$ contains $(p,v)$ after 
Line~\ref{line:reg:swmrma:readAudit}. 

Otherwise, if a read completes in $H$, note that 
an audit returns $\auditResult$ 
(Line~\ref{line:reg:swmrma:AuditReturn}), 
which contains the contents of all registers 
($R_a$ Line~\ref{line:reg:swmrma:AuditAddRead}).
Therefore, $(p,v)$ is in the response of every audit that 
starts after $\op{r}$ completes.
\end{proof}

\begin{lemma}\label{lem:reg:swmrma:Accuracy}
The set of pairs $\mathcal{P}$ returned by an audit
satisfies the Accuracy property.
\end{lemma}

\begin{proof}
If $(p_i,v)$ is in the set $\mathcal{P}$ returned by an audit,
then $(p_i,v)$ is in $\auditResult$ (Line~\ref{line:reg:swmrma:AuditReturn}).
Note that $\auditResult$ contains values read from registers $R_a$ 
(Line~\ref{line:reg:swmrma:AuditAddRead}),
and a reader $p_i$ writes the set of all values it read to $R_a[i]$.
Furthermore, if there is a pending read invoked by process $p_i$, 
such that $\mathcal{P}$ includes $(p_i,v)$ in its response, 
and that no read returned $v$ to process $p_i$ in $H'$, 
we complete the pending read by providing the response $v_i$ to $p$.
Thus, there is a read $\op{r}$ in $H'$ that returns $v$ to $p_i$.
\end{proof}

\begin{theorem} \label{theorem:reg:swmrma:correctness}
Algorithm~\ref{alg:AuditableAtomicRegister:swmrma} implements 
a single-writer multi-reader atomic register 
with multi-auditor regular audit.
\end{theorem}

We remark that this algorithm can be specialized to get 
\emph{single}-writer \emph{single}-reader registers,
and extended to get \emph{multi}-writer multi-reader atomic registers.
In both cases, the algorithm can provide regular audit for 
one or many auditors.


\section{Conclusion}\label{sec:summary}
This paper studies the synchronization power of auditing an 
atomic read / write register, in a shared memory system.
We consider two alternative definitions of the audit operation, 
one which is atomic relative to the read and write operations,
and another that is regular. 
The first definition is shown to have a strong synchronization power,
allowing to solve consensus; the number of processes that can solve
consensus corresponds to the number of processes that can audit the register. 
We also implement an atomic audit operation, 
using swap and fetch\&add for a single auditor, 
and compare\&swap for multiple auditors. 
On the other hand, the weaker, regular audit can be implemented 
from ordinary reads and writes. 

We studied single-writer registers and leave the interesting question 
of registers with multiple writers to future work.

In most practical systems supporting auditing, there is a single 
auditor (e.g. a trusted third party or the data owner), 
and it seems that a regular audit would suffice. 
Stronger forms of auditing are needed when users do not trust 
a third party auditor, so that auditing is performed 
in a distributed manner. 
Determining the precise requirements in practical systems is 
outside the scope of this paper, but our results indicate that 
if a too-strong notion is enforced, 
it would incur high synchronization cost.

\bibliography{references}

\clearpage
\appendix


\end{document}